\documentclass[fleqn,usenatbib]{mnras}

\usepackage{newtxtext,newtxmath}

\usepackage[T1]{fontenc}

\DeclareRobustCommand{\VAN}[3]{#2}
\let\VANthebibliography\thebibliography
\def\thebibliography{\DeclareRobustCommand{\VAN}[3]{##3}\VANthebibliography}

\usepackage{graphicx}	
\usepackage{amsmath}	

\graphicspath{{.}{figs/}}

\newcommand{\sigsfr}{\Sigma_{\rm SFR}}
\newcommand{\siggas}{\Sigma_{\rm gas}}
\newcommand{\ha}{{\rm H}\alpha}
\newcommand{\hb}{{\rm H}\beta}
\newcommand{\aco}{\alpha_\mathrm{CO}}
\newcommand{\ewha}{W$_\mathrm{H\alpha}$}

\title[Star-formation efficiency in early-type galaxies]{The resolved star-formation efficiency of early-type galaxies}

\author[T G. Williams, F. Belfiore et al.]{Thomas G. Williams,$^{1}$\thanks{E-mail: thomas.williams@physics.ox.ac.uk (TGW)}
Francesco Belfiore,$^{2}$
Martin Bureau,$^{1}$
Ashley T. Barnes,$^{3, 4}$
Frank Bigiel,$^{4}$
\newauthor
Woorak Choi,$^{5}$
Ryan Chown,$^{6}$
Dario Colombo,$^{4}$
Daniel A. Dale,$^{7}$
Timothy A. Davis,$^{8}$
Jacob Elford,$^{9}$
\newauthor
Jindra Gensior,$^{10}$
Simon~C.~O.~Glover,$^{11}$
Brent Groves,$^{12}$
Ralf S.\ Klessen,$^{11,13,14,15}$
Fu-Heng Liang,$^{1}$
\newauthor
Hsi-An Pan,$^{16}$
Ilaria Ruffa,$^{2, 8}$
Toshiki Saito,$^{17}$
Patricia S\'{a}nchez-Bl\'{a}zquez,$^{18, 19}$
Marc Sarzi$^{20}$
\newauthor
and Eva Schinnerer$^{21}$
\\
$^{1}$Sub-department of Astrophysics, Department of Physics, University of Oxford, Keble Road, Oxford OX1 3RH, UK\\
$^{2}$INAF — Osservatorio Astrofisico di Arcetri, Largo E. Fermi 5, I-50125, Florence, Italy\\
$^{3}$European Southern Observatory (ESO), Karl-Schwarzschild-Stra{\ss}e 2, 85748 Garching, Germany\\
$^{4}$Argelander-Institut f\"{u}r Astronomie, Universit\"{a}t Bonn, Auf dem H\"{u}gel 71, 53121 Bonn, Germany\\
$^{5}$Department of Physics \&\ Astronomy, McMaster University, 1280 Main St W, Hamilton, ON L8S 4M1, Canada\\
$^{6}$Department of Astronomy, The Ohio State University, 140 West 18th Ave, Columbus, OH 43210, USA\\
$^{7}$Department of Physics and Astronomy, University of Wyoming, Laramie, WY 82071, USA\\
$^{8}$Cardiff Hub for Astrophysics Research \&\ Technology, School of Physics \&\ Astronomy, Cardiff University, Queens Buildings, The Parade, Cardiff, CF24 3AA, UK\\
$^{9}$Instituto de Estudios Astrof\'{i}sicos, Facultad de Ingenier\'{i}a y Ciencias, Universidad Diego Portales, Av. Ej\'{e}rcito Libertador 441, Santiago, Chile\\
$^{10}$Institute for Astronomy, University of Edinburgh, Royal Observatory, Edinburgh EH9 3HJ, Scotland, United Kingdom\\
$^{11}$Universit\"{a}t Heidelberg, Zentrum für Astronomie, Institut f\"{u}r Theortische Astrophysik, Albert-Ueberle-Str. 2, 69120 Heidelberg, Germany\\
$^{12}$International Centre for Radio Astronomy Research, University of Western Australia, 7 Fairway, Crawley, 6009 WA, Australia\\
$^{13}${Universit\"{a}t Heidelberg, Interdisziplin\"{a}res Zentrum f\"{u}r Wissenschaftliches Rechnen, Im Neuenheimer Feld 225, 69120 Heidelberg, Germany}\\
$^{14}${Harvard-Smithsonian Center for Astrophysics, 60 Garden Street, Cambridge, MA 02138, USA}\\
$^{15}${Elizabeth S. and Richard M. Cashin Fellow at the Radcliffe Institute for Advanced Studies at Harvard University, 10 Garden Street, Cambridge, MA 02138, USA}\\
$^{16}$Department of Physics, Tamkang University, No.151, Yingzhuan Road, Tamsui District, New Taipei City 251301, Taiwan\\
$^{17}$Faculty of Global Interdisciplinary Science and Innovation, Shizuoka University, 836 Ohya, Suruga-ku, Shizuoka 422-8529, Japan\\
$^{18}$Instituto de F\'{i}sica de Part\'{i}culas y del Cosmos (IPARCOS), Universidad Complutense de Madrid, E-28040 Madrid, Spain\\
$^{19}$Departamento de F\'{i}sica de la Tierra y Astrof\'{i}sica, Universidad Complutense de Madrid, E-28040 Madrid, Spain\\
$^{20}$Armagh Observatory and Planetarium, College Hill, Armagh BT61 9DG, UK\\
$^{21}$Max Planck Institut f\"ur Astronomie, K\"onigstuhl 17, 69117 Heidelberg, Germany\\
}

\date{Accepted XXX. Received YYY; in original form ZZZ}

\pubyear{2024}

\begin{document}
\label{firstpage}
\pagerange{\pageref{firstpage}--\pageref{lastpage}}
\maketitle

\begin{abstract}
Understanding how and why star formation varies between galaxies is fundamental to our comprehension of galaxy evolution. In particular, the star-formation efficiency (SFE; star-formation rate or SFR per unit cold gas mass) has been shown to vary substantially both across and within galaxies. Early-type galaxies (ETGs) constitute an extreme case, as about a quarter have detectable molecular gas reservoirs but little to no detectable star formation. In this work, we present a spatially-resolved view of the SFE in ten ETGs, combining state-of-the-art Atacama Large Millimeter/submillimeter Array (ALMA) and Multi Unit Spectroscopic Explorer (MUSE) observations. Optical spectroscopic line diagnostics are used to identify the ionized emission regions dominated by star-formation, and reject regions where the ionization arises primarily from other sources. We identify very few regions where the ionization is consistent with pure star formation. Using $\ha$ as our SFR tracer, we find that previous integrated measurements of the star-formation rate based on UV and 22\micron\ emission are systematically higher than the SFR measured from $\ha$. However, for the small number of regions where ionization is primarily associated with star formation, the SFEs are around 0.4~dex higher than those measured in star-forming galaxies at a similar spatial resolution (with depletion times ranging from $10^8$ to $10^{10}$~yr). Whilst the SFE of ETGs is overall low, we find that the SFEs of individual regions within ETGs can be similar to, or higher than, similar sized regions within star-forming galaxies.
\end{abstract}




\section{Introduction}\label{sec:intro}

\begin{figure*}
    \includegraphics[width=\textwidth]{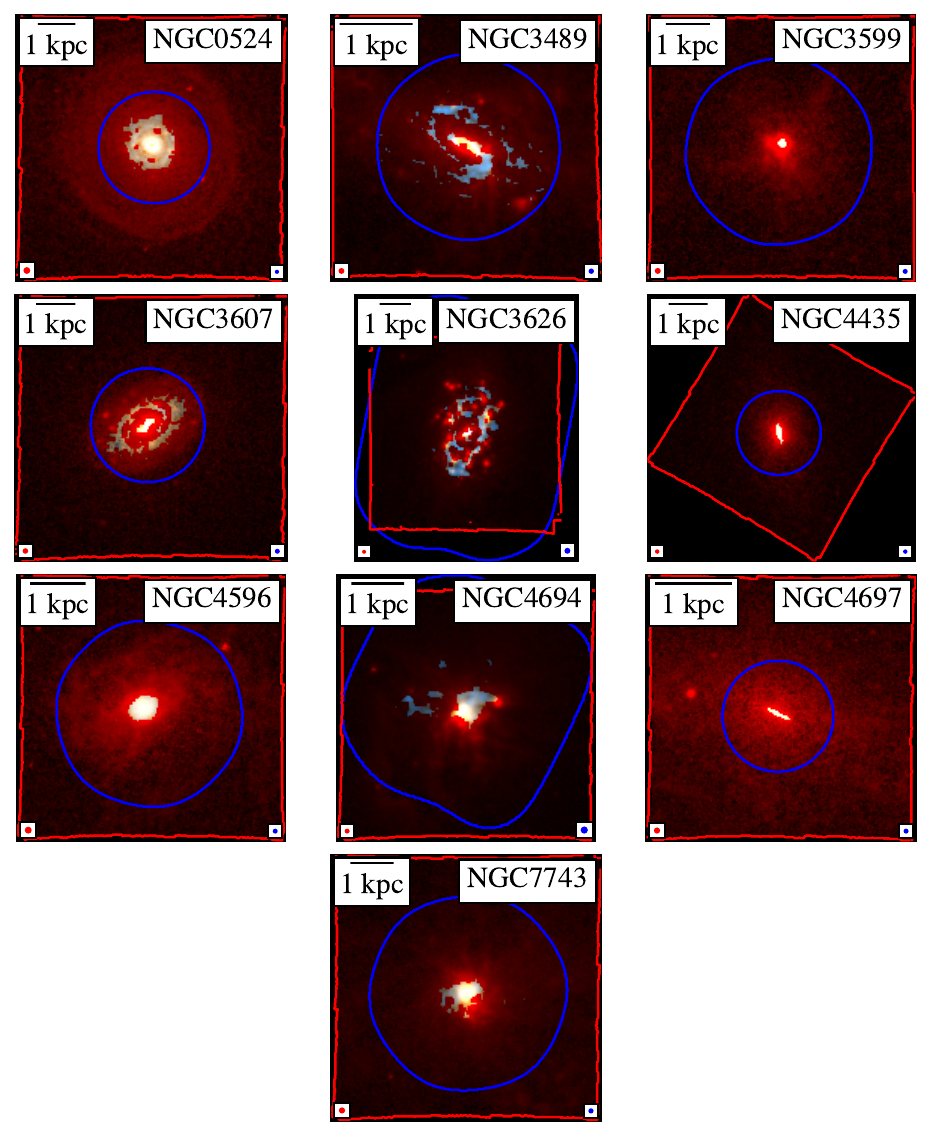}
    \caption{MUSE $\ha$ (red) and ALMA CO (blue) data of our sample of ten ETGs. The field-of-view of each observation is shown as a contour in the same colour. The data are shown at their native spatial resolutions, with the MUSE point-spread function as a red circle in the bottom-left and the ALMA synthesised beam as a blue circle in the bottom-right of each panel. A 1~kpc scalebar is also shown in the top-left of each panel.}
    \label{fig:data_overview}
\end{figure*}

The processes driving the conversion of molecular gas into stars are fundamental to shape the Universe we see today. However, the dominant factors setting these processes are not well understood. Is the star-formation efficiency (SFE; the star-formation rate divided by the cold gas mass, or inverse of the gas depletion timescale) set by local, cloud-scale properties \citep[e.g.][]{2020Sun, 2021Rosolowsky}? Or, is this primarily shaped by large-scale processes such as the evolutionary state of the galaxy or features such as spiral arms and bars \citep[e.g.][]{2021Querejeta, 2021Villanueva}? To disentangle what drives star formation in galaxies, we require cloud-scale observations of both the molecular gas (the raw reservoir for star formation) and the star-formation rate (SFR). To fully sample the galactic parameter space, we require studies across the whole range of galaxy types \citep{2024SchinnererLeroy}.

Much work has gone into understanding the relationship between the SFR and the total amount of cold gas (i.e.\ atomic and molecular hydrogen) present in a galaxy, often referred to as the `star-formation' or `Kennicutt-Schmidt' (KS) relation \citep{1959Schmidt, 1998Kennicutt, 2019delosReyesKennicutt}. This relationship has been studied at both entire galaxy (i.e. integrated) scales \citep[e.g.][]{1998Kennicutt, 2019delosReyesKennicutt} and resolved ($\leq$kpc) scales \citep[e.g.][]{2008Bigiel, 2010Schruba, 2015Casasola, 2018Williams, 2022Abdurrouf, 2022SanchezGarcia, 2024Zanchettin}. Across much of the star-forming galaxy population, there is a tight power-law relationship (with a power-law index of $\approx$1.5) between the integrated surface density of SFR ($\sigsfr$) and the surface mass density of (total cold) gas ($\siggas$). This relationship may hold across cosmic time \citep{2013Freundlich}. On entire galaxy scales, deviations from the KS relation (both in terms of the power-law index, and a $\approx$0.3~dex offset from the relation) exist for starburst galaxies and low-metallicity dwarfs \citep{2012KennicuttEvans}, which may indicate that star formation proceeds somewhat differently in these types of objects. With spatially-resolved ($\sim$kpc) observations, \cite{2008Bigiel} showed that the molecular gas is the more important element driving star formation (with a KS power-law index of almost unity). Pushing to even higher spatial resolutions, at scales of about 100~pc, the intrinsic scatter of the KS relation increases by $\approx$0.1~dex \citep{2010Schruba, 2010Onodera, 2018Williams}. At these scales, different galactic features (e.g. spiral arms, bars) occupy different regions of the KS relation \citep{2021Pessa}.

Until now, however, most efforts have been focused on star-forming, spiral galaxies, with comparatively little attention given to the more quiescent early-type galaxies (ETGs). Contrary to the classical picture of these galaxies being `red-and-dead', about a quarter of them harbour molecular gas \citep{2011Young, 2019Davis, 2019Ruffa, 2023Ruffa}, and many host low levels of star formation \citep{2014Davis, 2024RuffaDavis}. Unlike spiral galaxies, however, this molecular gas is typically confined to the central $\approx$1~kpc of these galaxies \citep{2024RuffaDavis}.  For a sample of 43 ETGs, \cite{2007Combes} reported SFEs similar to those of spiral galaxies.
\cite{2014Davis}, however, reported that the SFEs of ETGs are typically suppressed by a factor $\approx$ 2.5 on the KS relation, which they attribute to the extreme dynamical conditions in the centres of these galaxies stabilising the gas against collapse, in turn inhibiting star formation \citep{2022Lu, 2024Lu}. However, the limited spatial resolution of these studies did not resolve the star formation coincident with this centrally-concentrated molecular gas, and so it is unclear whether the entire molecular gas reservoir is star forming. If the entire area covered by star-forming regions is more compact than the molecular gas in unresolved observations, this could naturally lead to galaxies being suppressed on the KS relation on integrated scales, whilst the SFE of the resolved star-forming regions may be closer to that of the KS relation. Indeed, \cite{2024Lu} showed that, with resolved observations, individual star-forming regions lie much closer to the KS relation in the ETG NGC~0524. A similar result was also found in the ETG NGC~5128 (Cen A) by \cite{2019Espada}.

In this work, we have studied the SFE in a spatially resolved manner across a sample of 10 ETGs (Hubble $T$\textless0) using a combination of Very Large Telescope (VLT) Multi Unit Spectroscopic Explorer (VLT-MUSE) and Atacama Large Millimeter/submillimeter Array (ALMA) data. The layout of this work is as follows. In Section \ref{sec:data}, we present an overview of the data and associated reductions. In Section \ref{sec:results} we present our local measurements of the SFE, comparing to similar resolution observations in main-sequence, star-forming galaxies. In Section \ref{sec:discussion} we discuss our results, before concluding in Section \ref{sec:conclusions}.

\section{Data}\label{sec:data}

\begin{table*}
	\centering
	\caption{Overview of the data used in this study.}
        \label{tab:gal_info}
	\begin{tabular}{cccccccc}
		\hline
		Galaxy & Distance$^{\rm a}$ & $T$$^{\rm b}$ & $\log_{10}M_\star/M_\odot$ & $\log_{10}$(SFR$/{\rm M}_\odot~{\rm yr}^{-1}$)  & $\log_{10}$(SFR$/{\rm M}_\odot~{\rm yr}^{-1}$)  & MUSE Native Res. & ALMA Native Res. \\
            & Mpc & & $z$0MGS$^{\rm c}$ & $z$0MGS$^{\rm c}$ & This work$^{\rm d}$ & \arcsec & \arcsec \\
            \hline
            NGC~0524 & 23.3 & $-1.2\pm0.6$ & $11.0\pm0.1$ & $-0.6\pm0.2$ & $-3.8<x<-1.5$ & 1.08 & 0.46 \\
            NGC~3489 & 11.9 & $-1.2\pm0.9$ & $10.2\pm0.1$ & $-1.2\pm0.2$ & $-3.9<x<-1.6$ & 0.87 & 0.75 \\
            NGC~3599 & 19.9 & $-2.0\pm0.5$ & $10.0\pm0.1$ & $-1.2\pm0.2$ & $-4.1<x<-1.8$ & 0.89 & 0.66 \\
            NGC~3607 & 22.2 & $-3.2\pm1.4$ & $11.1\pm0.1$ & $-0.5\pm0.2$ & $-3.5<x<-1.1$ & 0.79 & 0.65 \\
            NGC~3626 & 20.1 & $-0.8\pm1.1$ & $10.4\pm0.1$ & $-0.6\pm0.2$ & $-1.9<x<-1.0$ & 0.57 & 1.22  \\
            NGC~4435 & 16.7 & $-2.1\pm0.5$ & $10.4\pm0.1$ & $-0.8\pm0.2$ & $-5.0<x<-1.3$ & 0.71 & 0.68 \\
            NGC~4596 & 15.8 & $-0.8\pm0.8$ & $10.6\pm0.1$ & $-1.0\pm0.2$ & $-4.6<x<-2.3$ & 1.11 & 0.65 \\
            NGC~4694 & 15.8 & $-1.8\pm0.7$ & $9.94\pm0.1$ & $-0.9\pm0.2$ & $-1.3<x<-1.2$ & 0.69 & 1.22 \\
            NGC~4697 & 11.4 & $-4.5\pm0.8$ & $10.8\pm0.1$ & $-1.1\pm0.2$ & $-\infty<x<-2.4$ & 0.93 & 0.66 \\
            NGC~7743 & 20.3 & $-0.9\pm0.9$ & $10.3\pm0.1$ & $-0.7\pm0.2$ & $-5.3<x<-1.0$ & 0.93 & 0.70 \\
            \hline
	\end{tabular}
        \\Notes: (a) \citet{2017Steer}. (b) Numerical Hubble type \citep{2014Makarov}. (c) \citet{2019Leroy}. The uncertainties are dominated by calibration uncertainty and listed as 0.1~dex for the stellar mass and 0.2~dex for the SFR. (d) We calculate an upper and a lower limit to the SFR based on BPT diagnostics. See Section \ref{sec:ha_sfr}.
\end{table*}

In this Section, we describe the data processing steps adopted for our MUSE and ALMA data. For homogeneity, we work at a fixed spatial resolution of 150~pc, the common worst resolution of this dataset (given the various galaxy distances and angular resolutions of the MUSE and ALMA data). We provide an overview of these data in Figure \ref{fig:data_overview}, and list the important parameters of the sample galaxies in Table \ref{tab:gal_info}.

\subsection{MUSE}\label{sec:data_muse}

The MUSE data used here are part of the Physics at High Angular Resolution in Nearby Galaxies \citep[PHANGS;][]{2021Leroy} survey, and the galaxies are part of the MUSE (PHANGS-MUSE) extended sample, which will be described in detail in a forthcoming publication from the PHANGS team. These targets have been selected as the complete sample of nearby ($D$~\textless~25~Mpc) ETGs that already have high angular (\textless1\arcsec) resolution ALMA observations, visible from the VLT. These galaxies well sample the range of stellar masses and morphologies seen in the overall ETG population \citep{2022Davis}. We summarise here a few key properties of the observations and data reduction, especially where they differ from the properties of the main PHANGS-MUSE sample described in \cite{2022Emsellem}.

MUSE data covering an approximate wavelength range of 480--680~nm with a mean spectral resolution of $R\approx3000$ were obtained via program 109.2332.001 (PI: Belfiore) for all targets except NGC~3489 and NGC~3626, for which we used archival data (0104.B-0404(A), PI: Erwin). We remove NGC~1317 and NGC~4457 from the sample observed in program 109.2332.001, as both of these galaxies are classified as spirals in the NASA/IPAC Extragalactic Database (NED). Each galaxy was targeted using  one $1\arcmin\times1\arcmin$ pointing in the wide-field mode of the instrument. Each pointing was observed for a total of 2440~s, divided into four exposures, except for NGC~3489 and NGC~3626 for which the total integration time was 1760~s. Furthermore, the data for NGC~3489 and NGC~3626 were obtained using the ground-layer adaptive optics correction, while the other datasets were observed in seeing-limited mode. These observations were generally carried out using the PHANGS-MUSE observing setup, leading to a typical $3\sigma$ flux sensitivity in the $\ha$ line of around $5\times10^{-19}~{\rm erg~s^{-1}~cm^{-2}}$ per 0.2\arcsec\ spaxel.

The data were reduced using  {\tt pymusepipe},\footnote{\url{https://pypi.org/project/pymusepipe/}} a customised Python wrapper to the MUSE data reduction pipeline \citep{2020Weilbacher} developed by the PHANGS team \citep{2022Emsellem}. {\tt pymusepipe} offers additional functionalities to the MUSE data reduction pipeline, including tools for absolute astrometric alignment and for fixing the absolute flux calibration to an external image. For our targets, images from the Sloan Digital Sky Survey \citep[SDSS;][]{2000York} data release 13 \citep{2017Albareti} and proprietary data from the Dupont telescope were used as external anchors. Fully-reduced, astrometrically-corrected and flux-calibrated datacubes covering the entire spectral range were thus produced with the `native' spatial resolution (which varies from 0.57\arcsec\ to 1.11\arcsec; see Table \ref{tab:gal_info}).

The native-resolution cubes were convolved to a common spatial resolution using a two-step procedure. We first calculated the point-spread function (PSF) of the MUSE data by comparing a synthetic $r$-band image generated from the cube to the SDSS image of the same field. The PSF of the SDSS image was generated using the {\tt photo} software,\footnote{See \url{https://www.sdss4.org/dr17/imaging/images/\#psf} for details.} and we reconstructed the PSF using the provided parameters of the particular SDSS tile used. We then used a cross-convolution method to estimate the MUSE PSF, following \cite{2017Bacon}.\footnote{Using this package: \url{https://github.com/cloud182/musepsf}.} Once the PSF is calculated, parameterised as a  Moffat profile, we  homogenised the cube to the nearest Gaussian PSF to produce a `convolved, optimised' \citep[referred to as ``copt'', see][]{2022Emsellem} cube. We then convolved each cube slice with a Gaussian kernel to produce cubes at a fixed spatial resolution (and subsequently pixel size) of 150 pc. This means that we increase the signal-to-noise (S/N) via smoothing before any fitting is performed on the cube, rather than binning the fitted maps afterwards. 

Once our homogeneous cubes were produced, we created maps of emission line fluxes and kinematics using the MUSE Data Analysis Pipeline\footnote{\url{https://gitlab.com/francbelf/ifu-pipeline/}} described in \cite{2022Emsellem}. In short, the pipeline performs spectral fitting using a set of extended Medium resolution Isaac Newton Telescope (INT) Library of Empirical Spectra (eMILES) simple stellar population models \citep{2016Vazdekis}, simultaneously fitting the emission lines with a set of Gaussian functions. The fit is performed with the penalised pixel fitting (pPXF) Python routine described in \cite{2004CappellariEmsellem} and \cite{2017Cappellari}. The DAP also corrects for foreground Milky Way (MW) extinction. In this work, we use $\ha$\ as our tracer of the SFR. We correct this for internal extinction via the Balmer decrement, using H$\beta$. However, as described in Section \ref{sec:ha_sfr}, $\ha$\ emission can arise from a variety of sources, and so we also require a number of other lines to disentangle the dominant ionization source -- for this, we use various [N{\sc ii}], [S{\sc ii}], and [O{\sc iii}] lines, which are also covered by the MUSE observations.

\subsection{ALMA}\label{sec:data_alma}

Some of the ALMA data used here have previously been processed as described in \cite{2023Williams} using the PHANGS-ALMA pipeline \citep{2021Leroy}, and we refer readers to those works for a more detailed description. These are observations for the galaxies NGC~0524, NGC~3607, NGC~4435 and NGC~4697 (Programme IDs 2015.1.00466.S, 2015.1.00598.S, 2016.2.00053.S and 2017.1.00391.S). We also reduce data from Programme IDs 2018.1.00484.S, 2017.1.00766.S, 2017.1.00886.L and 2019.1.01305.S, which cover the galaxies NGC~3489, NGC~3599, NGC~3626, NGC~4596, NGC~4694 and NGC~7743. Briefly, we generated calibrated datasets using the {\tt scriptForPI} files provided by the observatory. All of our measurement sets contain a high-spectral resolution window for the line observations (typically around 1~km~s$^{-1}$), as well as a number of continuum spectral windows with coarser resolution. Using all available spectral windows, we performed a $uv$-continuum subtraction before combining all available calibrated data of each galaxy into a single dataset. This always includes high-resolution 12m data, as well as 7m data for all but NGC~4697. Total power (TP) data is also available for the galaxies NGC~3489, NGC~3599, NGC~3626, NGC~4596, NGC~4694 and NGC~7743. In this case, we image the TP data with the PHANGS-ALMA pipeline, and feather it in with the imaged interferometric cube. These observing setups are designed such that the extent of the CO is contained within the largest angular scale (LAS) of the observing configuration, so we expect flux recovery to be complete. 

For each galaxy and thus dataset, we carried out a shallow multi-scale {\tt CLEAN} \citep{2008Cornwell} to a depth of four times the root mean square (RMS) noise, before a deeper single-scale {\tt CLEAN} using the \cite{1974Hogbom} algorithm to the RMS noise. We use Briggs weighting with a robust parameter of 0.5, the same as the PHANGS-ALMA processing. We produced cubes at a spectral resolution of around 2.5~km~s$^{-1}$, using integer binning of the native spectral resolution. The synthesised beam was then circularised, and any blank space around the cube was removed to minimise space requirements. Our final ``native'' resolutions for each cube are given in Table \ref{tab:gal_info}, and range from 0.46\arcsec\ to 1.22\arcsec. As a final step, we convolved the cube to a fixed spatial resolution of 150~pc before creating moment maps, so that we can make direct comparisons with the MUSE data. We opt to convolve the cubes rather than taper the $uv$ data for direct comparison to the PHANGS data. As these two approaches typically agree well \citep[with differences $<10\%$;][]{2022Davis}, we are confident our chosen method does not bias our results. To ensure we have high confidence in isolating real emission in our intensity maps (at the cost of slightly decreased completeness), we calculated a `strict mask' to apply when producing each integrated intensity map. For details of the masking procedure, we refer readers to \cite{2006RosolowskyLeroy} and \cite{2021Leroy}.

\section{Results}\label{sec:results}

\subsection{What drives $\ha$ emission?}\label{sec:ha_sfr}

\begin{figure*}
    \includegraphics[width=\textwidth]{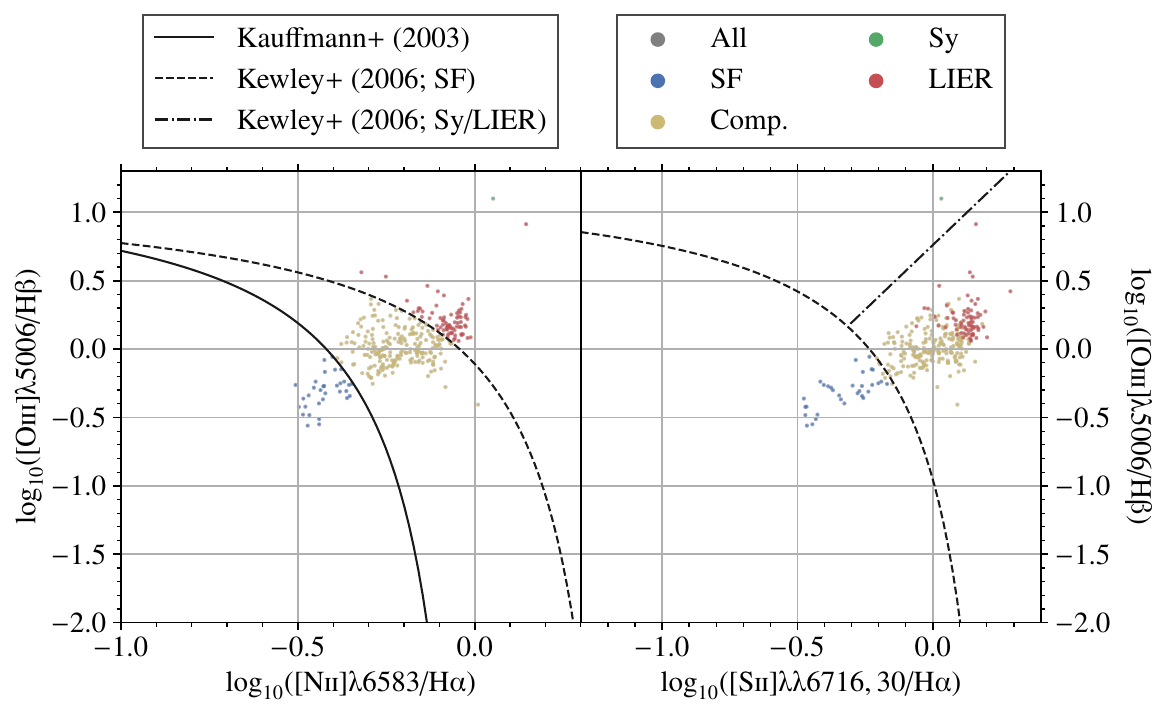}
    \caption{BPT diagram of individual spaxels within the galaxy NGC~4694. {\it Left}: [O{\sc iii}] -- [N{\sc ii}] diagnostic. {\it Right:}[O{\sc iii}] -- [S{\sc ii}] diagnostic. The demarcation from \citet{2003Kauffmann} is shown as a solid black line, and those from \citet{2006Kewley} as black dashed and dot-dashed lines. Data points are coloured by their positions in the BPT diagram, and are grey if they do not meet our signal-to-noise thresholds. The error-bars are typically on the order of 0.01 for these ratios as the lines are strongly detected, so these uncertainties are negligible.}
    \label{fig:ngc4694_bpt}
\end{figure*}

\begin{figure*}
    \includegraphics[width=\textwidth]{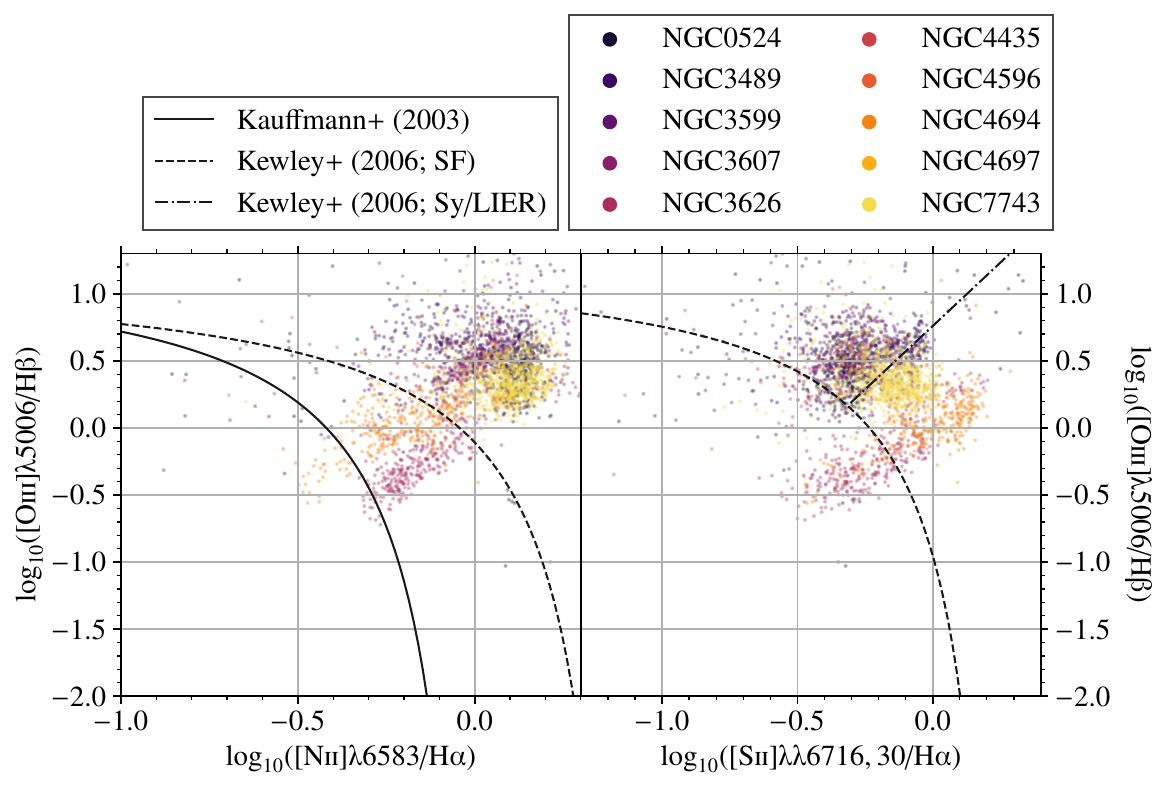}
    \caption{As Figure \ref{fig:ngc4694_bpt}, but for all the galaxies of our sample. In this case, the points are coloured by galaxy.}
    \label{fig:all_bpt}
\end{figure*}

\begin{figure*}
    \includegraphics[width=\textwidth]{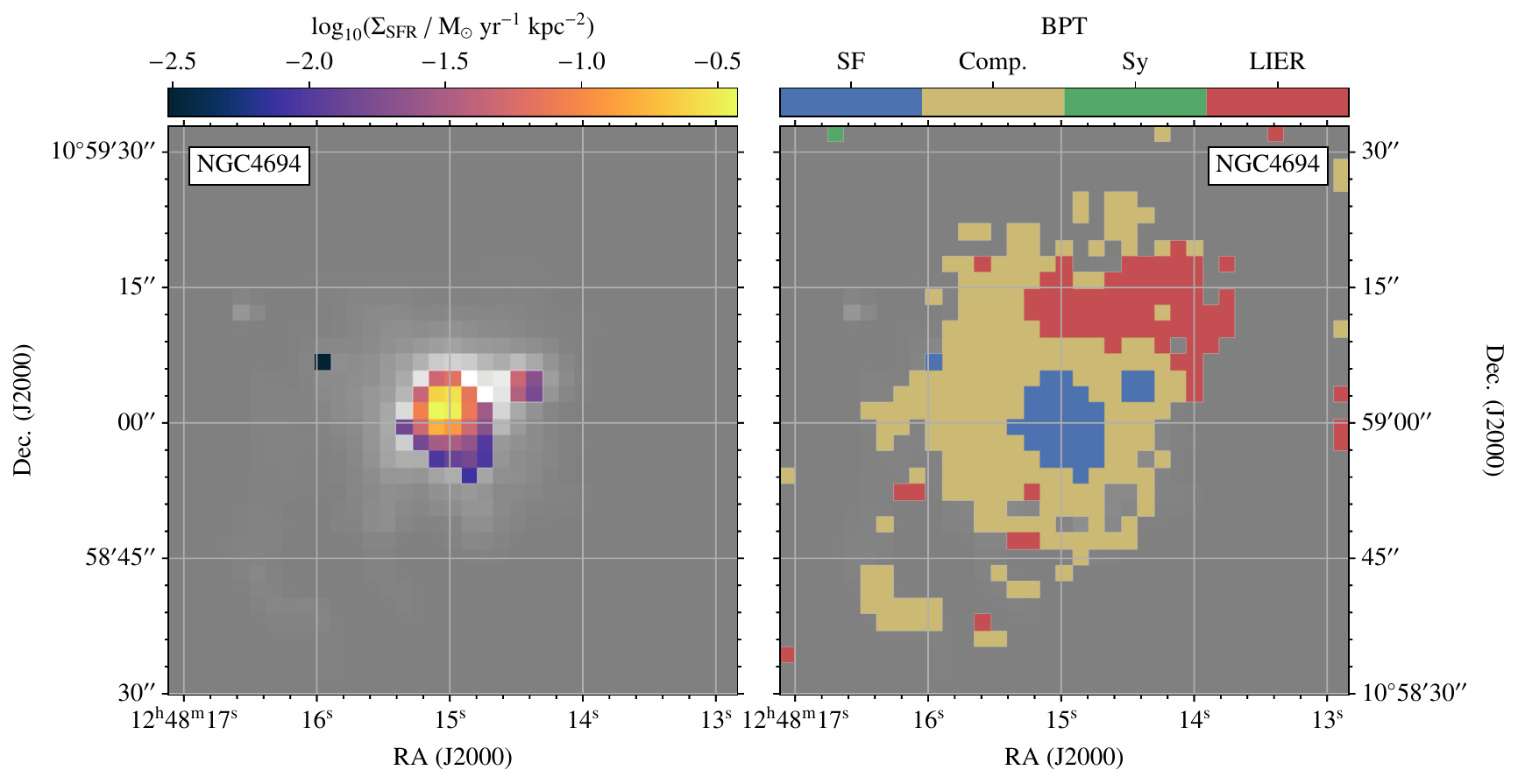}
    \caption{SFR (left) and BPT (right) maps of the galaxy NGC~4694. The pixels here correspond to a 150~pc-size region, and pixels are only coloured if they pass our S/N cuts, and the SFR maps are calculated only for spaxels that fall in the SF region of the BPT diagram. The greyscale shows $\ha$ emission in both cases. In the left panel, for the small number of spaxels classified as star-forming, we show the SFR surface densities in colour. The right panel shows spaxels coloured by their BPT classification, with the same colours as in Figure \ref{fig:ngc4694_bpt}. For this galaxy, the majority of the spaxels are defined as having their ionization dominated by processes other than star-formation.}
    \label{fig:ngc4694_sfr_map}
\end{figure*}

\begin{figure*}
    \includegraphics[width=\textwidth]{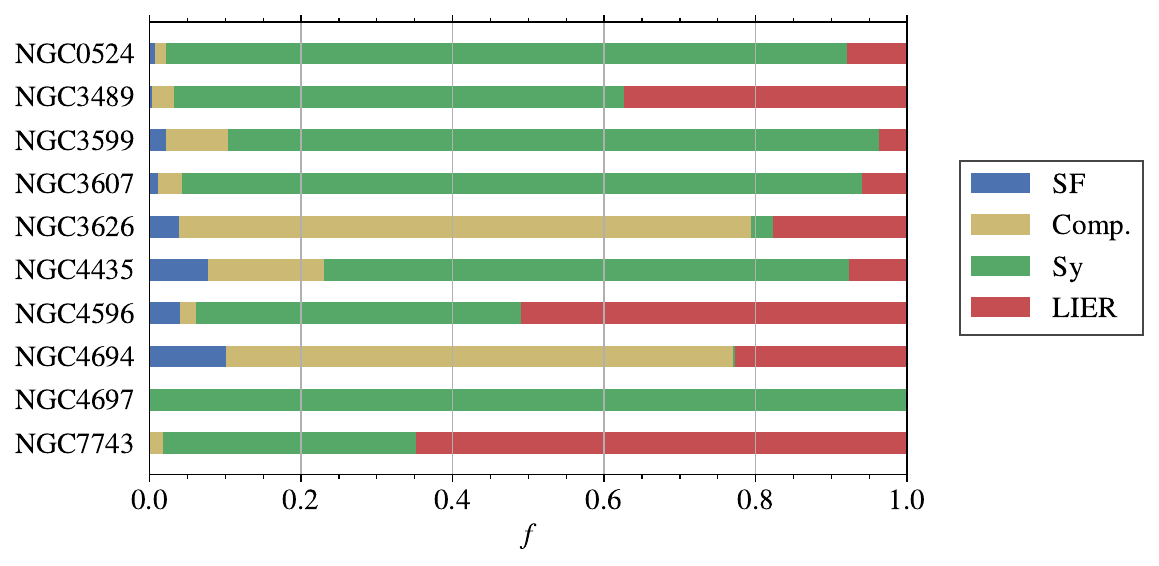}
    \caption{BPT classification of all the galaxies in our sample. Each bar is normalised to the number of spaxels where all relevant lines are securely detected in the map (i.e. non-detections due to low S/N are not included). The vast majority of the spaxels across all of our ETGs are not classified as dominated by star-formation.}
    \label{fig:bpt_barchart}
\end{figure*}

One of the primary benefits of optical integral-field spectroscopic (IFS) observations is that we can use optical diagnostics in the spectra not just to to classify the dominant sources of ionization, but to localise the source of this emission. The most common diagnostic for this is the \cite{1981Baldwin} diagram (BPT), which separates $\ha$ emission from star-forming regions from that of active galactic nuclei (AGN) or other sources of ionization such as supernova remnants and/or planetary nebulae \citep[e.g.][]{2022Santoro}. We separate these regions using the \cite{2003Kauffmann} and \cite{2006Kewley} line ratios, which use a combination of the [O{\sc iii}]/$\hb$, [N{\sc ii}]/$\ha$ and [S{\sc ii}]/$\ha$ ratios. We show an example of this for NGC~4694 in Figure \ref{fig:ngc4694_bpt} and the full sample in Figure \ref{fig:all_bpt}. Here, we use a S/N ratio cut of 3$\sigma$ for all the strong lines ($\ha$, H$\beta$, [N{\sc ii}], [S{\sc ii}] and [O{\sc iii}]) used to define that the emission has been well-detected (coloured points in these Figs, which we refer to as ``well classified''), the same as adopted by the PHANGS-MUSE team \citep{2022Santoro}. By binning spaxels to a fixed spatial resolution, the lines are either all detected with high significance or not detected at all, so the exact S/N cut does not affect our results. Figure \ref{fig:all_bpt} clearly shows that the majority of the data points do not lie in the purely star-forming regime of the BPT diagram, despite the widespread $\ha$ emission. We also show this classification as a map in Figure \ref{fig:ngc4694_sfr_map} for NGC~4694, where the majority of the well-classified spaxels in the BPT diagram are identified as AGN-like. Only a few spaxels are purely star forming across the entire sample, and are typically found further out in the disc (rather than in the galaxy centre). We note here that there may be some low level of star-formation in regions with low $\ha$\ luminosity,
and given that there are many such spaxels, this could form a significant contribution to the total SFR. We explore this further in Section \ref{sec:integrated_comparison}.

We stress here that spaxels not in the star-formation region of the BPT are not necessarily devoid of star-formation; rather, their ionization and thus emission is dominated by other sources, such as, for example, AGN \citep[e.g.][]{2006Kewley}, post-AGB (pAGB) stars \citep[e.g.][]{2010Sarzi}, or shocks \citep[e.g.][]{2004Groves}. In such cases, the $\ha$ fluxes should be considered upper limits on the ``true'' SFRs. However, given the low equivalent width of $\ha$ (see Sect. \ref{sec:whan}), the amount of star-formation in each individual region is likely to be low. Attempting to disentangle the relative contributions to the ionization from various mechanisms is beyond the scope of this paper, and we will simply treat these measurements as upper limits given the measured fluxes. We present analogous figures to Figure \ref{fig:ngc4694_bpt} and Figure \ref{fig:ngc4694_sfr_map} in Appendix \ref{app:all_bpts} for all of our sample galaxies.

We show the BPT classifications of all the spaxels of our galaxies in Figure \ref{fig:bpt_barchart}. Maximally, about 10~per~cent of all the spaxels are classified as having an ionizing radiation field which is dominated by star formation, with all of our sample galaxies having less than 10~per~cent of spaxels classified as star-forming. We note that occasionally very few spaxels are detected above our S/N threshold, and so a high percentage does not necessarily mean a large number of spaxels have been classified within a particular category. This highlights that star formation within ETGs is not the dominant source of ionization except in a few, small regions within each galaxy. In Section \ref{sec:whan}, we explore an alternative to the BPT diagnostics, but our findings are broadly similar to those obtained using the BPT classification -- the majority of emission regions within ETGs are not classified as purely star-forming.

\subsection{Comparison of SFRs to integrated measurements}\label{sec:integrated_comparison}

\begin{figure*}
    \includegraphics[width=\textwidth]{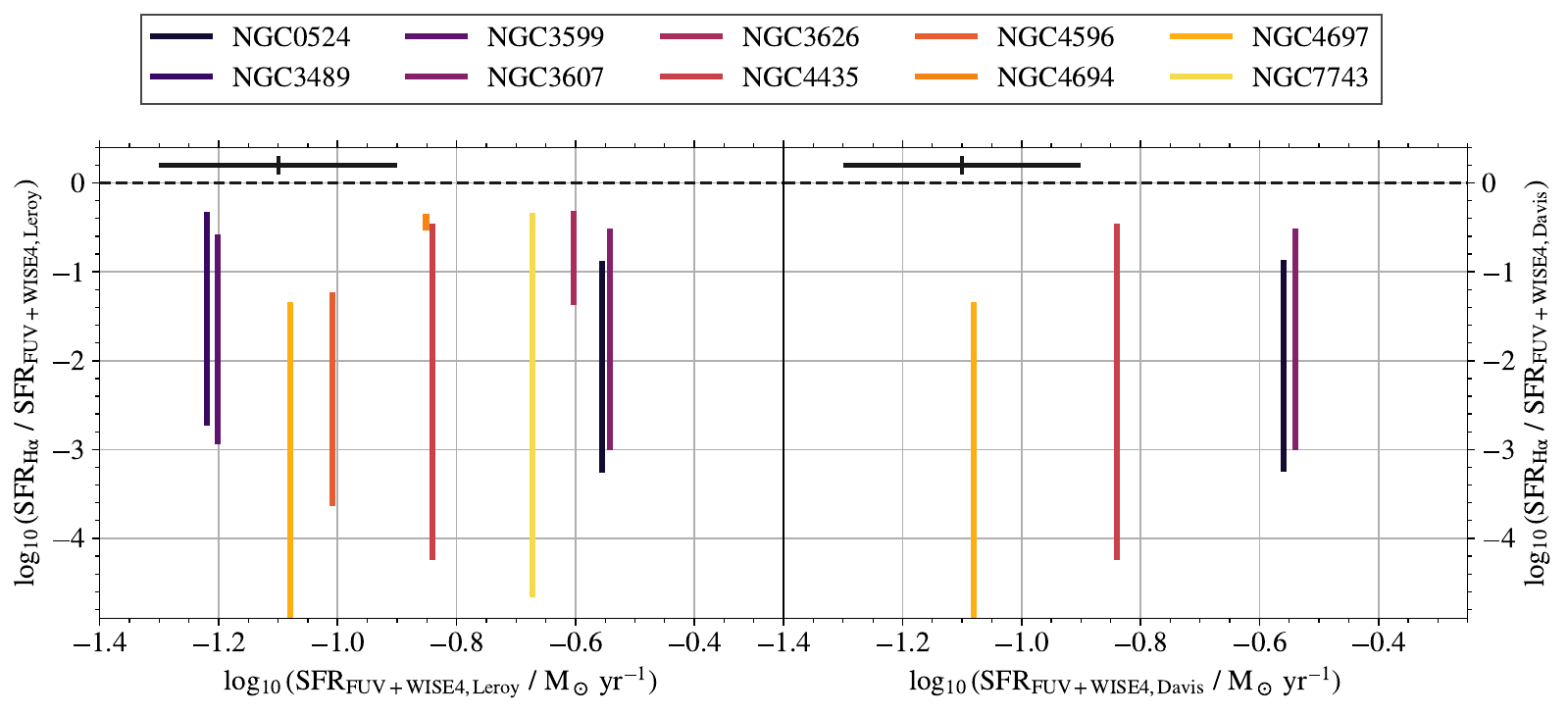}
    \caption{{\it Left}: Ratios of FUV+22\micron\ emission-based SFRs \citep{2022Leroy} to our $\ha$-derived SFRs as a function of the FUV+22\micron\ SFRs. The upper limit of each bar is calculated using the sum of all $\ha$ flux in the image, the lower limit using the sum of all $\ha$ flux in the star-forming region of the BPT diagram only. For NGC~4697, the lower limit is $-\infty$ (i.e.\ we find no star-forming spaxels within the the galaxy). The dashed line shows the 1:1 relation, and a representative error bar is shown in the top left (here, the $y$-axis error bar is the typical error from the $\ha$ SFR only). The FUV+22\micron\ SFRs are clearly higher than the $\ha$ SFRs, by more than one order of magnitude. {\it Right}: As in the left panel, but for the SFR values calculated by \citet{2014Davis, 2016Davis}.}
    \label{fig:sfr_int_resolve_comparison}
\end{figure*}

Throughout this work, we have used $\ha$ emission as our nominal tracer of the SFR. $\ha$ emission is an extremely common tracer for the SFR, as it arises due to ionization by young, massive stars \citep[e.g.][]{1983Kennicutt, 2021Pessa}. We note that at our spatial resolution, there may be biases that arise from incomplete initial mass function (IMF) sampling \citep{2014daSilva}. We expect the IMF to be fully sampled above an SFR of around $10^{-3}~{\rm M}_\odot~{\rm yr}^{-1}$ \citep[e.g.][]{2024Hu}, which may not be the case within individual 150~pc spaxels. Correcting for this incomplete IMF sampling is beyond the scope of this work. We also do not externally verify these $\ha$-based SFRs against an additional SFR tracer such as star-counting or radio continuum \citep[e.g.][]{2016Nyland} -- the former is impossible given the distances to our galaxies, and the latter will be the focus of future work. 

To transform an $\ha$ luminosity to a SFR, we first correct the emission for internal (galactic) extinction based on the Balmer ratio. For typical case B recombination \citep{1989Osterbrock}, and using a temperature of $10^4~{\rm K}$ and electron density $n_e$ of $100~{\rm cm}^{-3}$, the Balmer ratio is $\ha/\hb = 2.86$. The corrected $\ha$ flux is therefore
\begin{equation}\label{eq:balmer_correct}
    \ha_{\rm corr} = \ha_{\rm obs}\left(\frac{\ha_{\rm obs} / \hb_{\rm obs}}{2.86}\right)^{\frac{k_\alpha}{k_\beta - k_\alpha}},
\end{equation}
where the ``corr'' subscript indicates a corrected flux, the ``obs'' subscript indicates an observed flux, and $k_\alpha = 2.52$ and $k_\beta=3.66$, are the corresponding values of an \cite{1994ODonnell} extinction curve with an absolute to selective extinction, $R_V$ of $3.1$. We note that these are typical values, and specifically the ones adopted by \cite{2021Pessa}, to which we compare to later in our analysis.
To the Balmer-corrected $\ha$ fluxes, we apply the SFR prescription of \cite{2012KennicuttEvans}:
\begin{equation}\label{eq:ha_sfr}
    \log_{10}({\rm SFR}) = \log_{10}(\ha_{\rm corr}) - 41.27,
\end{equation}
where SFR is in $M_\odot~{\rm yr}^{-1}$ and the $\ha$ luminosity is in erg~s$^{-1}$. To compare our SFRs to integrated measures, we use the compilation of \cite{2019Leroy}, which uses a combination of Galaxy Evolution Explorer \citep[GALEX;][]{2005Martin} ultraviolet (UV) and Wide-field Infrared Survey Explorer \citep[WISE;][]{2010Wright} data (for all our targets, FUV and 22$\mu$m) emission to trace the SFRs. We do this to assess the importance of using optical diagnostics to identify true regions of star formation. We use as an upper limit the sum of all $\ha$ emission, and as a lower limit the sum of all $\ha$ emission in the purely star-forming region of the BPT diagram only. We show the results in the left panel of Figure \ref{fig:sfr_int_resolve_comparison}, demonstrating that the FUV+22\micron-derived SFRs are consistently higher than the $\ha$-based SFRs. When including all spaxels we obtain a higher value for the SFR, but still significantly lower than the integrated FUV+22\micron-derived SFRs. By removing emission from ionizing sources other than pure star formation, we obtain SFRs orders of magnitude lower. This is likely due to the FUV and/or 22\micron\ emission arising from an older stellar population 
(see \citealt{2017Viaene} for an example of 22\micron\ emission dominated by old stars). This has been discussed previously by \cite{2009Temi} and \cite{2014Davis}, where they attempted to correct for this older stellar population. Comparing to the SFR compilation in \cite[][which we show in the right panel of Fig. \ref{fig:sfr_int_resolve_comparison}]{2022Davis}, our measured upper SFRs are still significantly (often half an order of magnitude) below these corrected values. Therefore, blindly applying SFR prescriptions in these cases will lead to discrepant results; even when corrected, our results indicate these corrected values may still be significantly higher than the SFR as measured by $\ha$. Thus, the global SFRs of ETGs, which already appear suppressed relative to star-forming galaxies, may be even lower, given our $\ha$ SFRs.

\subsection{Star-formation efficiencies of ETGs}

\begin{figure*}
    \includegraphics[width=\textwidth]{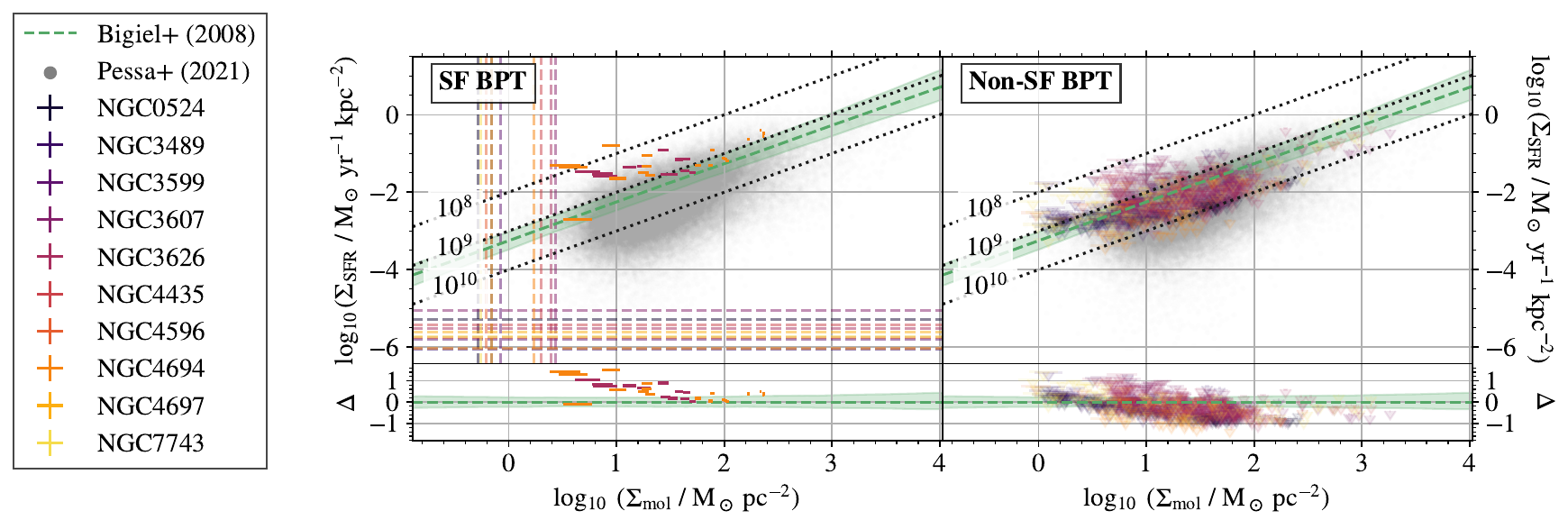}
    \caption{Resolved molecular Kennicutt-Schmidt relation of our sample ETGs. The left panel shows spaxels located within the star-forming region of the BPT diagram, the right panel spaxels outside of that region (formally SFR upper limits). These are shown as errorbars, but given the small error in the SFR surface density, they appear primarily as horizontal lines given the dynamic range of the plot. We show the relationship of \citet{2008Bigiel} as a green dashed line and the data points of \citet{2021Pessa} as a grey cloud. Lines of constant depletion times (labelled in years) are shown as black dotted lines, and the 3$\sigma$ completeness limit of each galaxy is shown as a coloured dashed (legend given on the left). The lower panels show the residual (in dex), having subtracted the \citet{2008Bigiel} relation.}
    \label{fig:etg_ks}
\end{figure*}

\begin{figure}
    \includegraphics[width=\columnwidth]{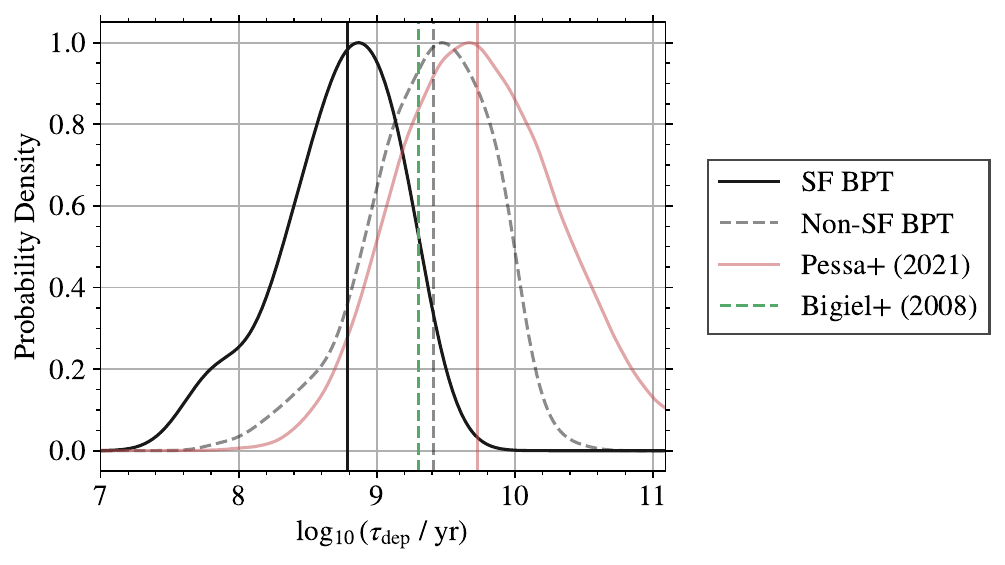}
    \caption{Kernel density estimate (KDE) plot showing the depletion times of spaxels within the star-forming region of the BPT diagram (black solid line) and non-star-forming region of the BPT diagram (grey dashed line; formally lower limits for the depletion time). We also show the distribution from \citet{2021Pessa} as a red line. For all these distributions, we calculate the KDE using bins of 0.01~dex. Vertical lines show the medians of these distributions, the green dashed line the median depletion time from \citet{2008Bigiel} and the red solid line the median depletion time from \citet{2021Pessa}. The depletion times are typically smaller than those of star-forming galaxies, measured at kpc and $\approx$100pc scale.}
    \label{fig:t_deps}
\end{figure}

To investigate the SFEs of our sample ETGs, we combine our MUSE and ALMA data to place these galaxies on the KS relation. We carry out this investigation at a fixed spatial resolution of 150~pc, which represents the best common physical resolution of the ALMA and MUSE maps (set by the worst seeing of the MUSE data). This resolution is sufficient to resolve, at least partially, individual star-forming regions \citep{2022Kim}, allowing us to study whether the overall low SFEs of our sample ETGs are due to widespread, but inefficient star-formation, or whether the star-forming regions occupy only a small area of the total regions containing molecular gas, but those regions form stars with a similar efficiency to spiral galaxies.

To convert a CO surface brightness to a molecular gas mass surface density, we use a CO conversion factor:
\begin{equation}\label{eq:co_conv}
    \alpha_\mathrm{CO(1-0)}=\frac{4.35}{R_{21}}\,\mathrm{M_\odot~pc^{-2}~\left(K~km~s^{-1}\right)}^{-1},
\end{equation}
where $R_{21}$ is the CO ({\it J}=2-1)/({\it J}=1-0) ratio of integrated fluxes expressed in brightness temperature units, and is here assumed to be $0.7$ \citep[a suitable ratio across a large range of galaxies; e.g.][]{2015Baldi, 2021DenBrok, 2022Leroy}. This factor also includes a correction for helium. We discuss this choice of conversion factor further in Section \ref{sec:alpha_co}. 

We show the resulting resolved KS plot in Figure \ref{fig:etg_ks}, with the measurements from star-formation-dominated spaxels in the left-hand panel, and the upper limits from spaxels in other regions of the BPT diagram on the right (these are upper limits on $\sigsfr$, and thus lower limits on the depletion timescale). We also show lines of constant depletion timescale (i.e.\ $\siggas$/$\sigsfr$) on both panels. We also show the analogous measurements at $\approx$100~pc resolution from PHANGS-MUSE \citep{2021Pessa} and the relationship for star-forming galaxies determined by \cite{2008Bigiel} which has been calculated at $\approx$1~kpc resolution\footnote{In \cite{2008Bigiel}, the best fit to their data does not include a molecular gas mass correction to account for helium ($\approx$36\%). We include this correction in the plots, which may at first appear inconsistent with this earlier work.}. Individual spaxels lie approximately within the scatter of the \cite{2021Pessa} data, but are elevated from the \cite{2008Bigiel} relation by a median of $0.44$~dex. Hence, whilst there are many regions where star formation dominates the ionisation within our galaxies, these few spaxels have SFEs higher than those of the local spiral galaxy population.

Figure \ref{fig:etg_ks} shows that spaxels we have defined as purely star-forming have elevated SFRs compared to the star-forming galaxy population. Taken at face value, this would mean that these spaxels have higher SFEs than typical spiral galaxies. However, this could also be due to the fact that these spaxels must have abnormally high SFRs for the ionization to be classified as purely star-forming. In this case, there could be a population of spaxels with SFEs close to or lower than in star-forming galaxies, but have the bulk of their ionization arising from processes other than star formation. Indeed, this is what the right panel of Figure \ref{fig:etg_ks} appears to show, with a median deviation from the \cite{2008Bigiel} relation of $-0.18$~dex. For this reason, we opt not to attempt to fit a power-law slope to these data points, as the fit would be both not meaningful, and ultimately misleading.

In Figure \ref{fig:t_deps}, we show the distributions of depletion times (inverses of SFEs) as a kernel density estimate plot (KDE). We use bins of 0.01~dex, and calculate the KDE using a \citet[][and also see this reference for an introduction to KDEs]{1986Silverman} bandwidth. We calculate this for spaxels lying in the star-forming and non-star-forming regions of the BPT diagram only. \cite{2008Bigiel} reported a typical depletion time of $2\times10^9$~yr, and for our purely star-forming spaxels, the average is around 0.5~dex lower ($6.2\times10^8$~yr). The average depletion time of star-forming regions is shorter in ETGs compared to star-forming galaxies. However, as in the discussion above, we may simply be selecting the tip of the iceberg, and biasing ourselves towards shorter depletion times by design. For the non-star-forming regions of the BPT diagram, we calculate a median depletion time of $2.5\times10^9$~yr, similar to the \cite{2008Bigiel} depletion time, calculated at kpc-scale. The median depletion time as calculated by \cite{2021Pessa} for 100~pc-scale regions is $5.3\times10^9$~yr. However, given that our SFRs are formally upper limits, the depletion times should be considered lower limits and could therefore truly be even higher. Considering the measurements from \cite{2021Pessa} at a similar spatial resolution to ours, while our depletion times are systematically smaller, they are lower limits and nevertheless still contained within the spread of measurements from the local spiral galaxy population.

To ensure our conclusions are not driven by the limited sensitivity of our data, we calculate a completeness limit for each galaxy. In the case of the CO, we measure the RMS in line-free channels of the cube, and assuming a typical linewidth of 10~km~s$^{-1}$, calculate the 3$\sigma$ completeness limit of a ``cloud'' (in reality, a PSF-sized spaxel in our study). For the SFRs, we use the median $\ha$\ error measured in each cube and convert this to a limiting SFR. Representative completeness limits are shown on the left panel of Figure \ref{fig:etg_ks} as coloured dashed lines. There could be significant numbers of regions with CO intensities below the detection threshold (as this is the limiting threshold in our study), for which we cannot measure an SFE. However, our completeness limits are below the majority of the PHANGS points, and thus our comparison to this sample should remain robust.

To conclude, the SFEs of the star-forming regions of ETGs may be similar to those of spiral galaxies on a spatially-resolved level. Whilst the SFEs for spaxels we define as having ionization dominated by star-formation are higher than the spiral galaxy population, by requiring this we bias ourselves to high SFEs by design. We find very few regions where the emission is consistent with being primarily due to ionization from star-formation. These regions often have elevated SFEs compared to the star-forming population, but the majority of spaxels have lower SFEs, which are typically lower than those of star-forming galaxies.

\section{Discussion}\label{sec:discussion}

Our results show that even at a spatially-resolved scale, ETGs have very few sites where star formation is able to dominate the ionization. In this Section, we explore an alternative method of classifying the ionizing radiation sources, link the low level of star formation back to the gas conditions, and discuss the limitations of our measurements of the molecular gas masses.

\subsection{Alternative classification of ionizing radiation}\label{sec:whan}

\begin{figure*}
    \includegraphics[width=\textwidth]{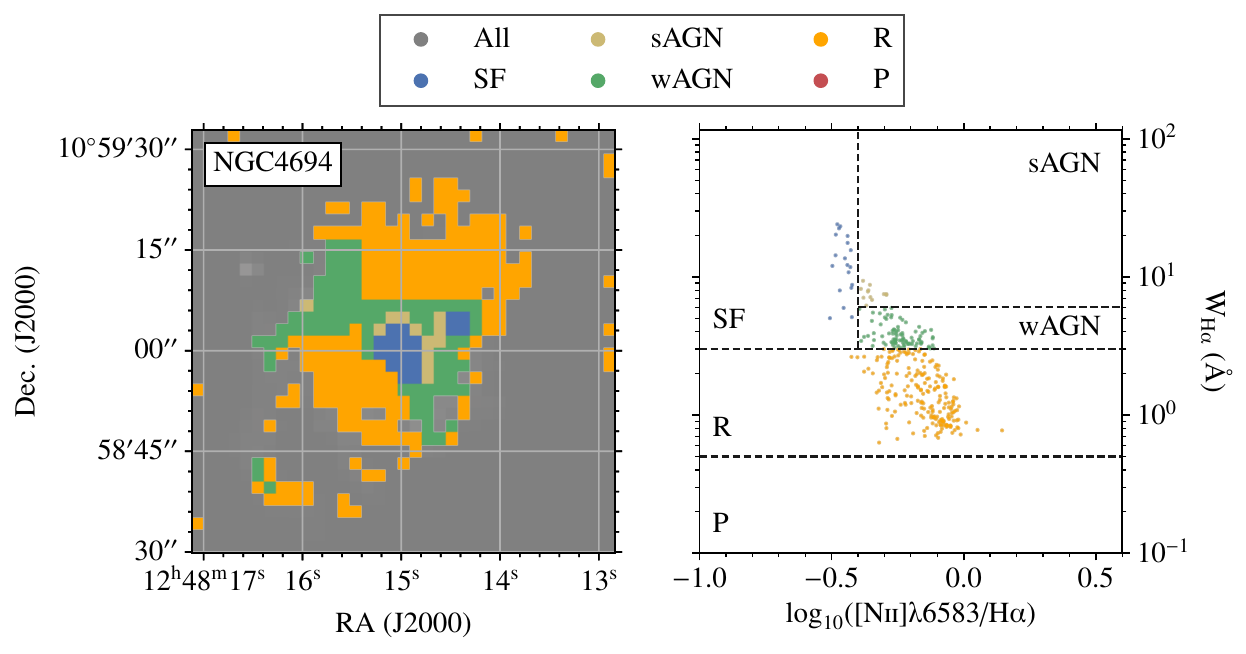}
    \caption{WHAN diagnostics of individual spaxels within the galaxy NGC~4694. {\it Left}: Map of the WHAN classes overlaid on an $\ha$ map of the galaxy. {\it Right}: WHAN classes defined using the \ewha -- ([N{\sc ii}]/$\ha$ plane. We use the classes of \citet{2011CidFernandes}: star-forming (SF), strong AGN (sAGN), weak AGN (wAGN), retired (R) and passive (P). Here, we have dropped the `galaxy' nomenclature from the retired and passive galaxies ued in \citet{2011CidFernandes}, to highlight that our observations are resolved.}
    \label{fig:ngc4694_whan}
\end{figure*}

\begin{figure}
    \includegraphics[width=\columnwidth]{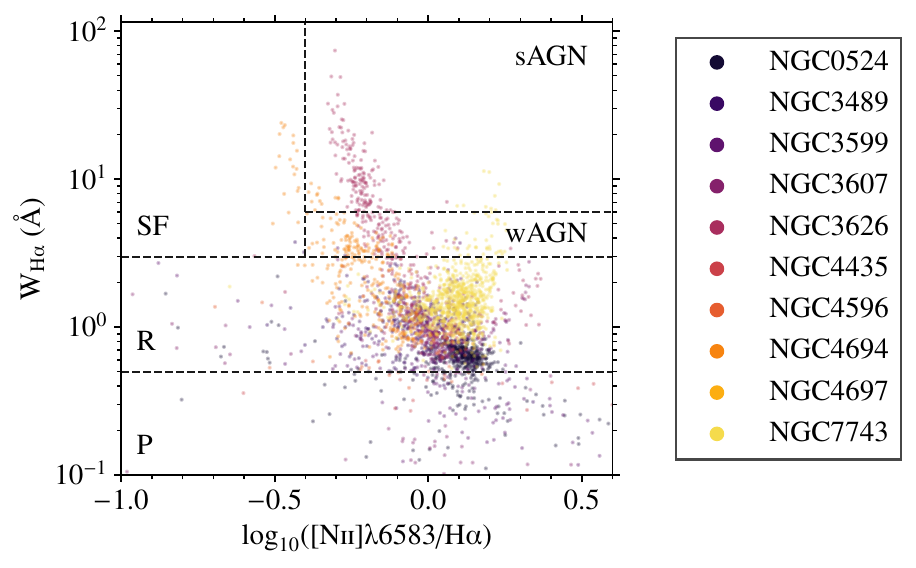}
    \caption{As Figure \ref{fig:ngc4694_whan}, but for all the galaxies of our sample. In this case, the points are coloured by galaxy.}
    \label{fig:all_whan}
\end{figure}

\begin{figure*}
    \includegraphics[width=\textwidth]{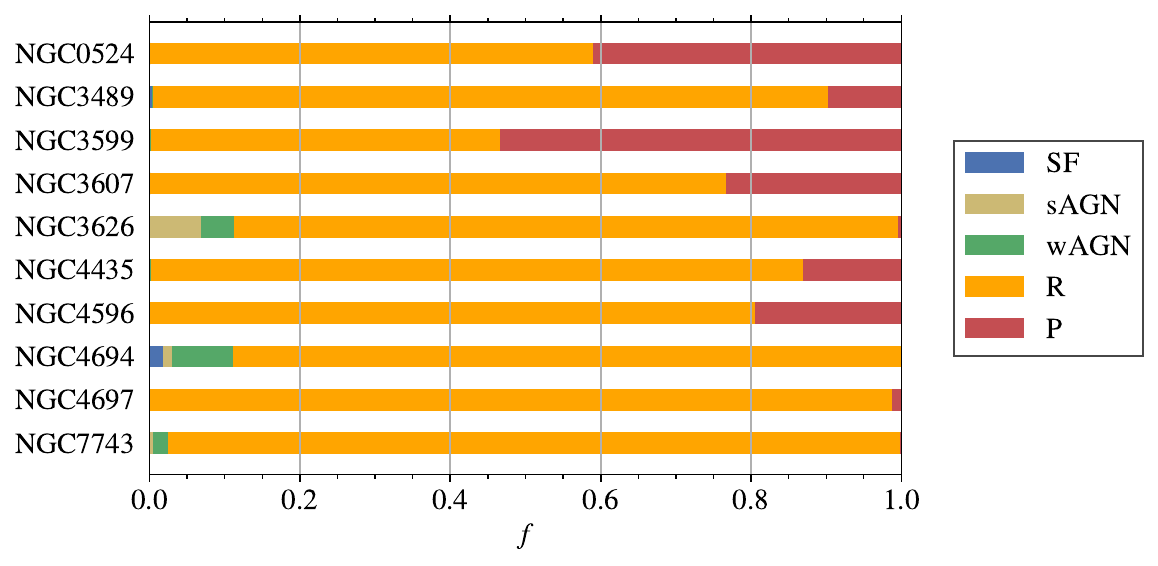}
    \caption{WHAN classification of all the galaxies in our sample. Each bar is normalised to the number of secure detections in the map (i.e. non-detections due to low S/N are not included). Much as for the BPT classification, most of the spaxels across all of our ETGs are not classified as star-forming.}
    \label{fig:whan_barchart}
\end{figure*}

Whilst the BPT diagram is popular for diagnosing sources of ionizing radiation, it requires the measurement of multiple emission lines which can be problematic in low S/N observations. As an alternative, \cite{2010CidFernandes} proposed a diagnostic based on the equivalent width of $\ha$ (\ewha) and the [N{\sc ii}]/$\ha$ ratio (the so-called WHAN diagram). This has the benefit of a reduced reliance on the often faint lines required for the BPT diagnostics, and removes the necessity of measuring multiple emission lines. This diagnostic was designed for integrated galaxy emission, and separates galaxies into SF, strong and weak AGN (sAGN and wAGN), `retired' galaxies heated by old stellar population, and `passive' galaxies which have no emission lines. We repeat our classifications of the spaxels within our ETGs using the classes advocated by \cite{2011CidFernandes}. To calculate W$_\mathrm{H\alpha}$, we follow the method outlined in \citet[their eqs. 11 and 17]{2019Westfall}. The results for NGC~4694 are shown in Figure \ref{fig:ngc4694_whan}. In this case, most of the spaxels lie in the `retired' category. This intuitively makes sense -- retired galaxies are primarily heated by an old stellar population, exactly what we expect for ETGs. The results for the full sample are shown in Figures \ref{fig:all_whan} and \ref{fig:whan_barchart}, and equivalents to Figure \ref{fig:ngc4694_whan} for all the galaxies of our sample in Appendix \ref{app:all_whans}. The picture is much the same as from the BPT classification -- a very small number of spaxels are classified as dominated by star-formation using the WHAN diagram. As with the BPT diagnostics, there may be some star formation hiding within these spaxels, but SF does not dominate the ionization budget.

\subsection{Gas conditions within ETGs}\label{sec:gas_conditions}

\begin{figure}
    \includegraphics[width=\columnwidth]{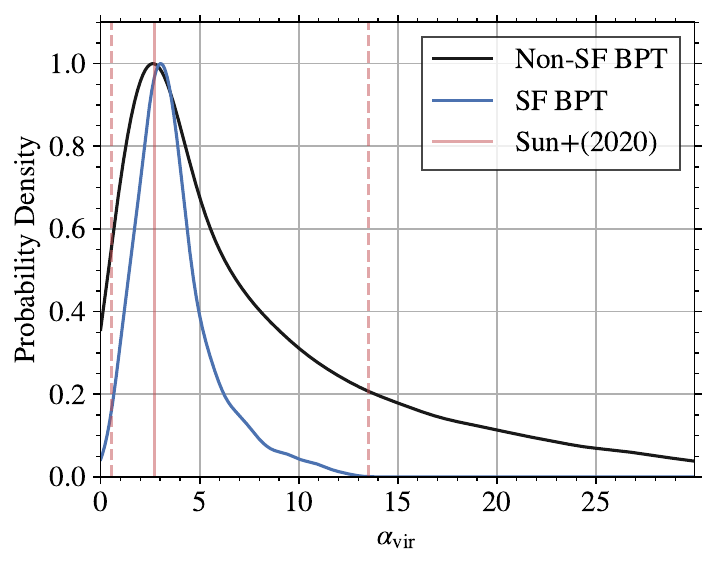}
    \caption{KDE plot showing the distribution of virial parameters of spaxels within the non-SF region of the BPT diagram (black) and spaxels within the star-forming regions of the BPT diagram (blue). We also include the area-weighted median and 68$^{\rm th}$ percentile spread reported for star-forming galaxies by \citet{2020Sun} as vertical red lines.}
    \label{fig:alpha_vir}
\end{figure}

\begin{figure*}
    \includegraphics[width=\textwidth]{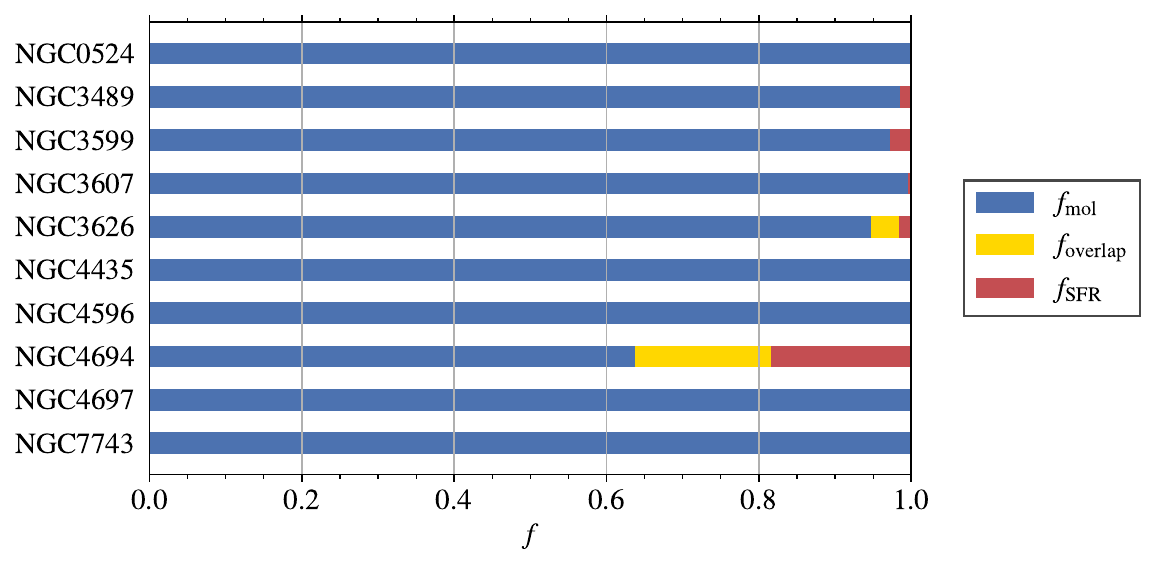}
    \caption{Fraction of CO-only spaxels (blue), SF-only spaxels (red; as defined by the BPT classification), and overlapping spaxels (yellow) of our sample ETGs. Only spaxels that are in both the ALMA and MUSE maps are included. CO is widespread across our sample ETGs, but star formation is very spatially-confined.}
    \label{fig:overlap_barchart}
\end{figure*}

\begin{figure*}
    \includegraphics[width=\textwidth]{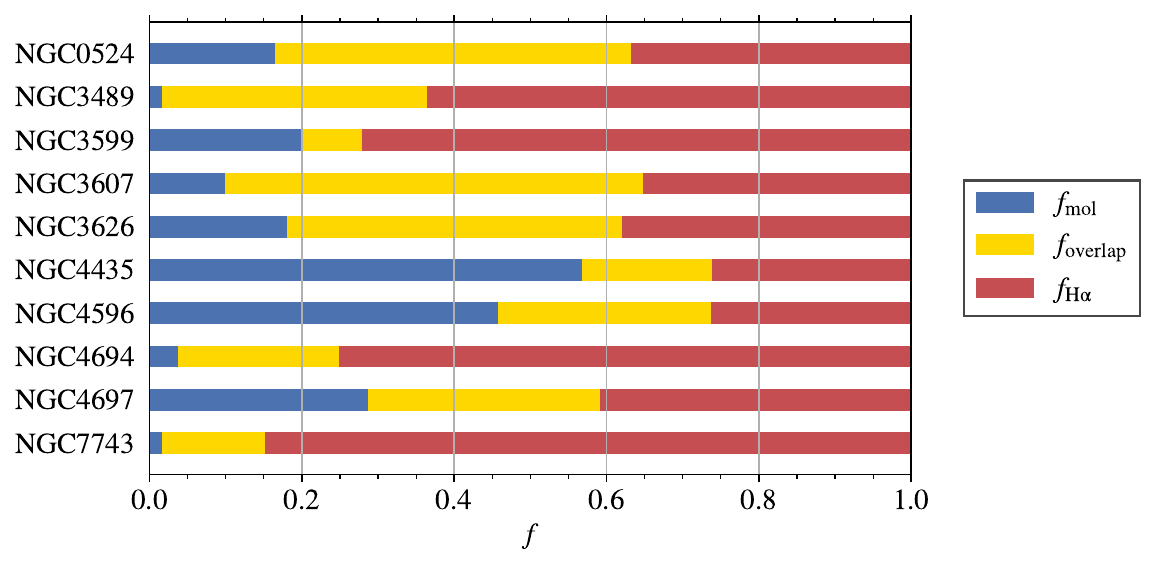}
    \caption{As Figure \ref{fig:overlap_barchart}, but instead using all detected $\ha$, rather than just spaxels within the SF region of the BPT diagram.}
    \label{fig:overlap_barchart_all_halpha}
\end{figure*}

\cite{2023Williams} showed that the molecular gas of ETGs is typically not in virial equilibrium (their sect. 3.2), and has turbulent pressures much higher than that of the star-forming galaxy population, more akin to the pressure within the central molecular zone (CMZ) of the MW (their sect. 3.3). We would expect both of these properties to suppress star formation, but in different ways. For the former, clouds are unbound and unlikely to coalesce over a long enough timescale for star formation to occur. For the latter, the increased turbulence is likely to lead to a reduced SFE \citep{2014Kruijssen}. Our results indicate the virial parameter may be the more pertinent quantity here, as there are very few sites where star-formation dominates the ionization, and these sites have SFEs comparable to those of late-type galaxies. To test this, we calculate the virial parameter \cite[][their eq. 9]{2023Williams} of each spaxel, accounting for the finite channel width of the ALMA data and assuming the spaxel size is the virial region size. We show the results in Figure \ref{fig:alpha_vir}, comparing spaxels within the non-star-forming region of the BPT diagram to those defined as star-forming in the BPT diagram. For the star-forming spaxels, $\alpha_{\rm vir}$ is on average somewhat smaller than the spaxels where ionization is dominated by non-SF sources, with a median of $3.3^{+2.2}_{-1.3}$ (errors here indicating the 16$^{\rm th}$ and 84$^{\rm th}$ percentiles) compared to the overall median of $5.5^{+10.5}_{-3.6}$, albeit with a significant spread. These are both somewhat larger than the value of unity (where the kinetic and gravitational potential energy balance), but all these spaxels have virial parameters similar to those typical of star-forming galaxies \citep{2020Sun}.

In Figure \ref{fig:overlap_barchart}, we show the fraction of spaxels that contain CO, lie in the SF region of the BPT diagram (i.e. ionized gas must be detected, and lies in the pure SF region of the BPT diagram), and their overlap. The picture here is extremely different from spiral galaxies \citep{2019Schinnerer}, where a significant amount of overlap between where CO and $\ha$\ is seen. The vast majority of spaxels only contain CO at our 150~pc resolution, despite this being comparable to the typical separation between star formation sites \citep{2022Kim}. Given the very small amount of overlap, rather than us catching CO before it forms stars, the majority of the molecular gas is unlikely to transform into stars at all (see also Fig. \ref{fig:alpha_vir}). This could be because large-scale dynamical processes, such as the gravitational potential of bulges and the effects of shear \citep[e.g.][]{2009Martig, 2020Gensior, 2021Gensior, 2021Liu, 2024Lu} keep the gas stable against collapse, allowing for widespread CO but only sparse and very localised SF in regions where the molecular material was able to overcome these forces and collapse \citep{2024RuffaDavis}.

In Figure \ref{fig:overlap_barchart_all_halpha}, we produce a similar bar chart to Figure \ref{fig:overlap_barchart} but instead including all spaxels that contain $\ha$, rather than just those that lie in the SF region of the BPT diagram. The fractions are significantly different to Figure \ref{fig:overlap_barchart}, as $\ha$\ emission is widespread across these galaxies meaning we have a much lower CO-only fraction, and a much higher overlap fraction. Comparing this to Figure \ref{fig:overlap_barchart}, it is clear that this ionized gas cannot predominantly originate from star-formation.

\begin{figure}
    \includegraphics[width=\columnwidth]{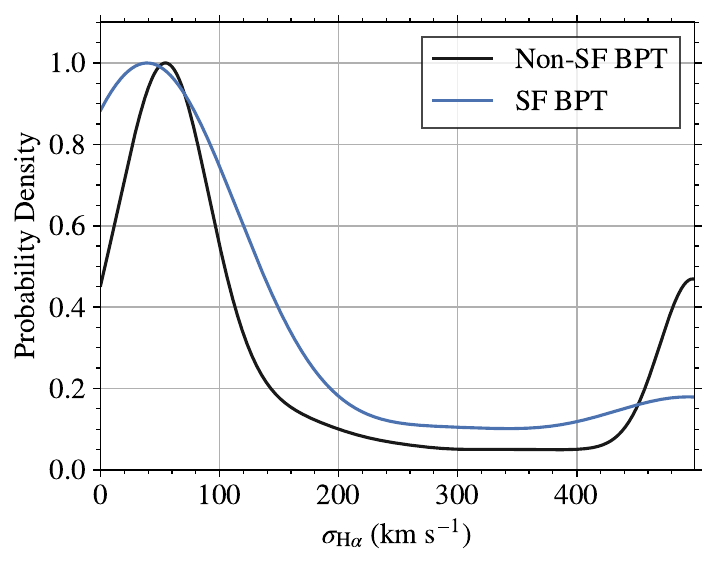}
    \caption{KDE plot showing the distribution of the $\ha$ velocity dispersions ($\sigma_{\rm H\alpha}$) of spaxels within the non-SF region of the BPT diagram (black) and spaxels within the star-forming regions of the BPT diagram (blue). The peak above 400~km~s$^{-1}$ are either poor fits or potential supernova remnants \citep{2022Santoro}.}
    \label{fig:sig_ha}
\end{figure}

We also expect the velocity dispersion of the $\ha$\ to be smaller in regions of more active star formation, as outside of these, $\ha$ emission arises from ionization by sources other than star-formation \citep[e.g.][]{2013Davis, 2020DellaBruna}, and stellar feedback leads to less turbulent regions than, for instance, shocks \citep[e.g.][]{2008Allen}. We thus calculate the intrinsic $\ha$ velocity dispersion ($\sigma_{\rm H\alpha}$) by subtracting in quadrature the MUSE line-spread function (LSF) from measured line-of-sight velocity distribution (LOSVD), to perform a deconvolution of the LSF from the LOSVD. We compare the distribution of the resulting velocity dispersions of star-forming spaxels (diagnosed using the BPT diagram) to the overall distribution across all galaxies in Figure \ref{fig:sig_ha}. The peak of the $\sigma_{\rm H\alpha}$ distribution of the star-forming spaxels lies at a slightly smaller dispersion than that of all of the spaxels within the maps (medians of 42 and 74~km~s$^{-1}$ respectively, albeit with a significant spread of around 15 and 40~km~s$^{-1}$, respectively), in agreement with earlier findings \citep{2013Davis}.

\subsection{The CO-to-H$_2$ conversion factor}\label{sec:alpha_co}

Throughout this work, we adopted a standard MW $\aco$ \citep[e.g.][]{2013Bolatto}. As our sample ETGs are likely to have gas-phase metallicities similar to that of the MW \citep[given their approximately MW-like mass and the relationship between mass and metallicity, e.g.][]{2004Tremonti}, we do not expect large variations of the CO-conversion factor, although we note that this conversion factor affects the measured molecular gas mass linearly, and therefore increasing $\aco$ by a factor of, for example, two will decrease the SFE by a factor of two. Particularly noteworthy is recent work by \cite{2024Teng}, who reported that $\aco$ is often smaller in galaxy centres (exactly the regions probed in this work), by up to an order of magnitude. If that were also the case for ETGs, then our SFE measurements would become correspondingly {\it higher}. However, as no direct measurement exists for $\aco$ within ETGs (although \citealt{2015Utomo} find a Galactic $\aco$ assuming virial equilibrium for the clouds within the ETG NGC~4526), we currently have no means to confirm the true CO-conversion factor of these galaxies.

\section{Conclusions}\label{sec:conclusions}

In this work, we combined deep, high-spatial resolution MUSE and ALMA data of a number of ETGs to study the SFEs of these `red-and-dead' galaxies. Using a BPT classification as well as WHAN diagnostics, very few (although not zero) regions within these galaxies are classified as having their ionization dominated by star formation, with the majority of ionization coming from AGN, pAGB stars, or shocks (BPT) or being classified as ``retired'' or ``passive'' (in the WHAN). These star-forming regions typically have smaller $\ha$ velocity dispersions than the surrounding gas, as well as smaller virial parameters. 

Comparing our measured SFRs to these earlier measurements, the FUV+22\micron\ SFR measurements are higher by at least half an order of magnitude and as much as four orders of magnitude in the worst case scenario with respect to our $\ha$ SFRs. Although the presence of an older stellar population can be corrected for \citep[e.g.][]{2014Davis}, we find that these corrected SFRs still appear too high compared to the $\ha$\ SFR. This work also highlights the potential power of optical spectroscopy, as we can use optical diagnostics to isolate and localise the sources of ionizing radiation -- which is clearly critical in the ETG case. However, given the binary nature of the BPT classification, regions with star formation that do not dominate the total ionization may be missed, leading to a bias towards low amounts of star formation.

Given the high spatial resolution (150~pc) of our observations, we also performed a resolved study of the star-formation (KS) relation of these galaxies. Regions dominated by ionisation from star-formation within our sample ETGs typically are elevated in SFR compared to similar regions in star-forming galaxies, with SFEs around 0.4~dex higher. However, these SFEs are likely biased high, as by requiring star formation to be the dominant ionization source, we may be restricted to regions with abnormally high SFEs. The low SFEs previously reported in the literature based on integrated measurements \citep[e.g.][]{2014Davis} appear to be due to efficient star formation being restricted to a small number of regions within these galaxies given our $\ha$ measurements, whilst the bulk of the gas forms stars at a very low level.

Overall, our results paint a picture of ETGs as hosting molecular gas, but with an overall lack of star formation. Regions where star formation dominates the ionization are few and far between, either because the majority of the molecular gas is unable to collapse into stars (due to the extreme dynamics in the centres of these galaxies), or because the low level of star formation only allows the most extreme ionised regions to be detected. In these detected regions, the efficiency at which star formation proceeds is similar to or higher than the typical SFEs in star-forming spiral galaxies. Further work to cross-check SFR measurements (with {\it JWST} narrow-band observations or radio continuum likely being the optimal choices) and anchor the CO-to-H$_2$ conversion factors offer promising and achievable avenues for further progress.

\section*{Acknowledgements}

The authors would like to thank the anonymous referee for their timely and constructive report, which helped to improve the paper. TGW would like to thank P. Woolford-Williams, for support during the preparation of this manuscript, and Milo Williams for excellent companionship. JG gratefully acknowledges funding via STFC grant ST/Y001133/1. SCOG acknowledges financial support from the European Research Council via the ERC Synergy Grant ``ECOGAL'' (project ID 855130) and from the Heidelberg Cluster of Excellence (EXC 2181 - 390900948) ``STRUCTURES'', funded by the German Excellence Strategy. RSK acknowledges financial support from the European Research Council via the  Synergy Grant ``ECOGAL'' (project ID 855130),  from the German Excellence Strategy via the Heidelberg Excellence Cluster  ``STRUCTURES'' (EXC 2181 - 390900948), and from the German Ministry for Economic Affairs and Climate Action in project ``MAINN'' (funding ID 50OO2206). RSK also thanks the 2024/25 Class of Radcliffe Fellows for their company and for highly interesting and stimulating discussions. HAP acknowledges support from the National Science and Technology Council of Taiwan under grant 113-2112-M-032-014-MY3. DC gratefully acknowledges the Collaborative Research
Center 1601 (SFB 1601 sub-project B3) funded by the Deutsche Forschungsgemeinschaft (DFG, German Research Foundation) –
500700252.

Based on observations collected at the European Southern Observatory under ESO programmes 0104.B-0404 (PI: Erwin), 0108.B-0215 (PI: Erwin) and 0109.B-0540 (PI: Belfiore). This paper makes use of the following ALMA data:\\
ADS/JAO.ALMA\#2015.1.00466.S,\\
ADS/JAO.ALMA\#2015.1.00598.S,\\
ADS/JAO.ALMA\#2016.2.00053.S,\\
ADS/JAO.ALMA\#2017.1.00391.S,\\
ADS/JAO.ALMA\#2017.1.00766.S,\\
ADS/JAO.ALMA\#2017.1.00886.L,\\
ADS/JAO.ALMA\#2018.1.00484.S and\\
ADS/JAO.ALMA\#2019.1.01305.S.\\
ALMA is a partnership of ESO (representing its member states), NSF (USA) and NINS (Japan), together with NRC (Canada), NSTC and ASIAA (Taiwan), and KASI (Republic of Korea), in cooperation with the Republic of Chile. The Joint ALMA Observatory is operated by ESO, AUI/NRAO and NAOJ. The NASA/IPAC Extragalactic Database (NED) is funded by the National Aeronautics and Space Administration and operated by the California Institute of Technology.

\section*{Data Availability}

Raw data is available via the ALMA (\url{https://almascience.nrao.edu/aq/}) and ESO (\url{https://archive.eso.org/eso/eso_archive_main.html}) archives. Maps and cubes are available at \url{https://www.canfar.net/storage/vault/list/AstroDataCitationDOI/CISTI.CANFAR/23.0016} for the data presented in \cite{2023Williams}, and \url{https://www.canfar.net/storage/vault/list/AstroDataCitationDOI/CISTI.CANFAR/25.0037} for new data in this work. 

Scripts underlying this study can be found at \url{https://github.com/thomaswilliamsastro/etgs_alma_muse_sfe}. PHANGS-ALMA pipeline keys to reproduce the ALMA reduction can be found at \url{https://github.com/thomaswilliamsastro/wisdom_alma_reduction}.



\bibliographystyle{mnras}
\bibliography{bibliography}



\appendix

\section{BPT diagrams for all galaxies}\label{app:all_bpts}

Here, we show figures analogous to Figure \ref{fig:ngc4694_bpt} and \ref{fig:ngc4694_sfr_map} for the other galaxies of our sample.

\begin{figure*}
    \includegraphics[width=\textwidth]{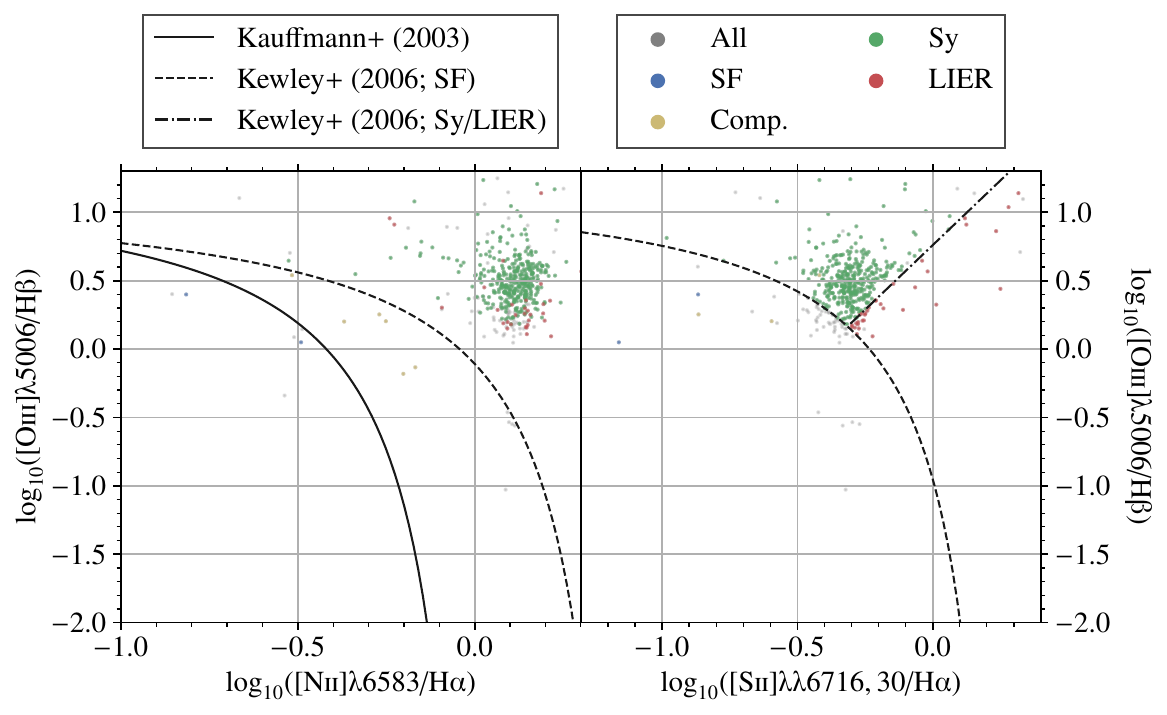}
    \caption{As Figure \ref{fig:ngc4694_bpt}, but for NGC~0524.}
    \label{fig:ngc0524_bpt}
\end{figure*}

\begin{figure*}
    \includegraphics[width=\textwidth]{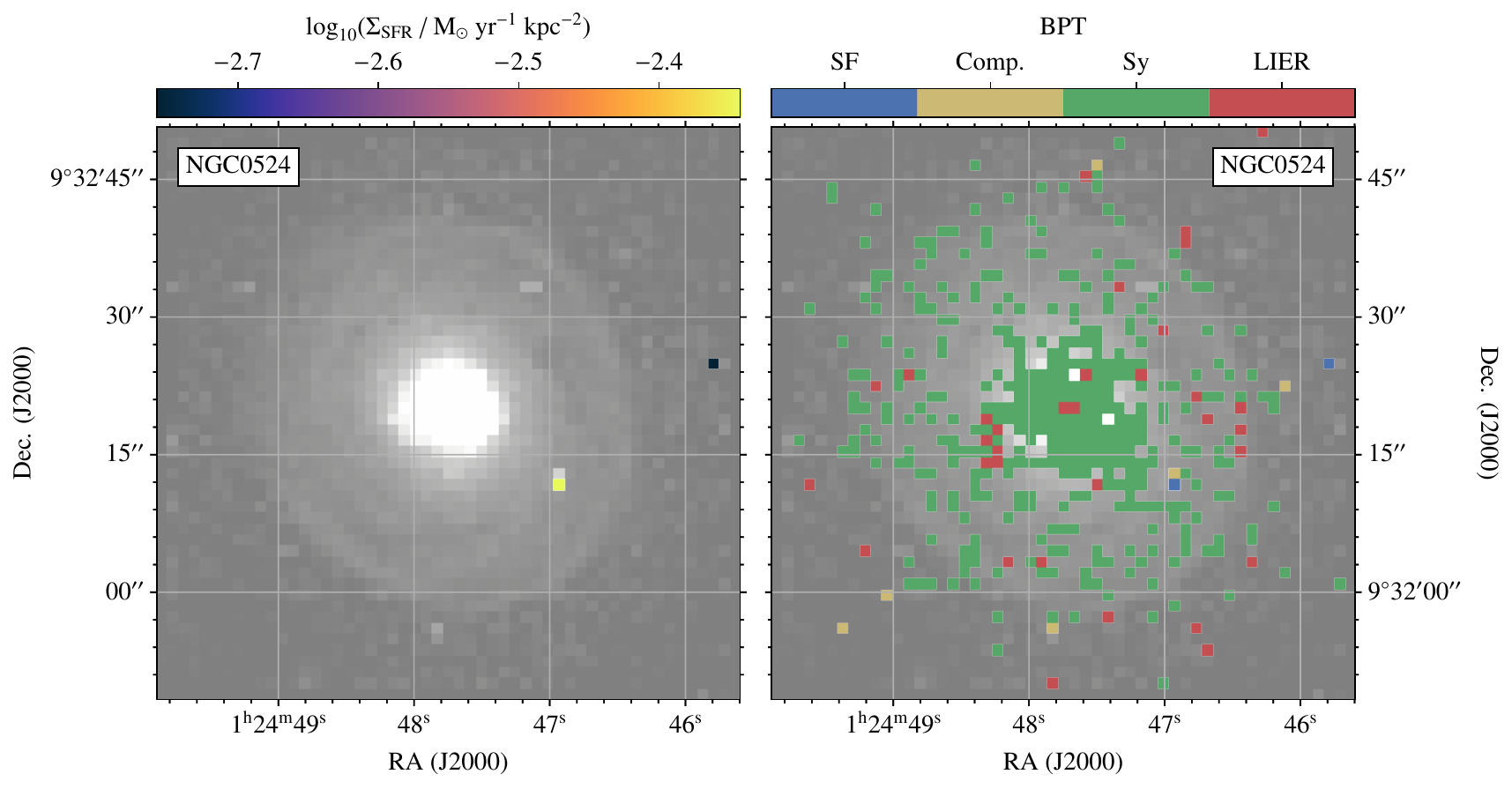}
    \caption{As Figure \ref{fig:ngc4694_sfr_map}, but for NGC~0524.}
    \label{fig:ngc0524_sfr_map}
\end{figure*}

\begin{figure*}
    \includegraphics[width=\textwidth]{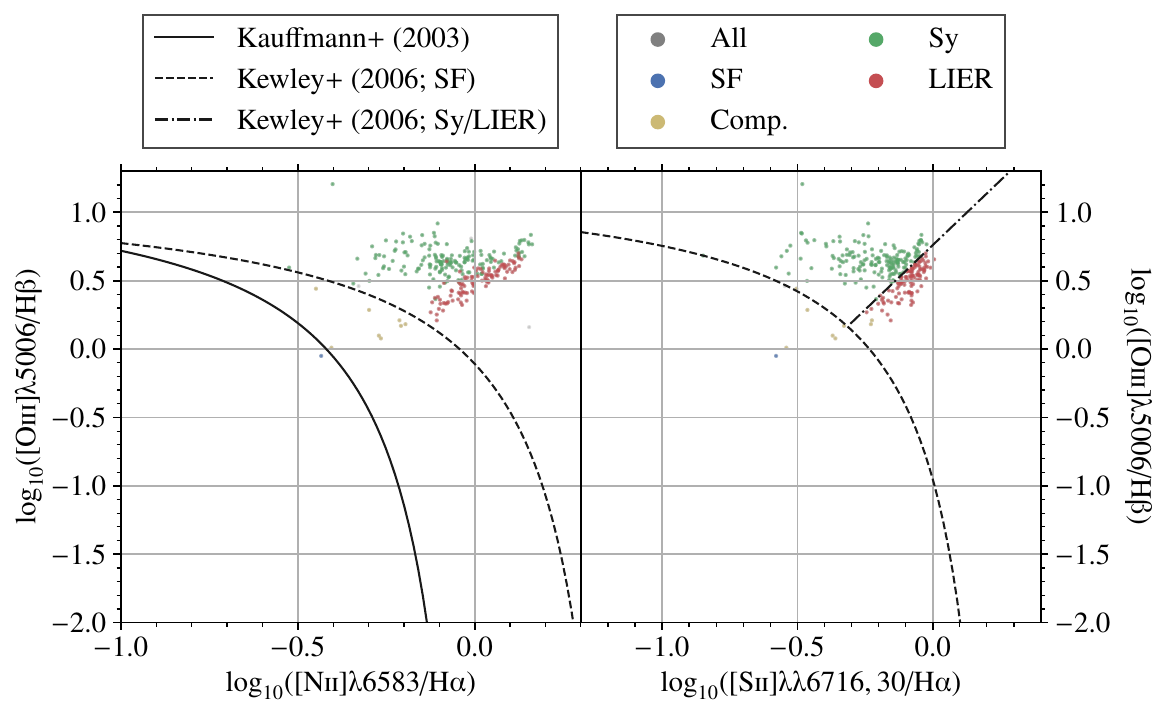}
    \caption{As Figure \ref{fig:ngc4694_bpt}, but for NGC~3489.}
    \label{fig:ngc3489_bpt}
\end{figure*}

\begin{figure*}
    \includegraphics[width=\textwidth]{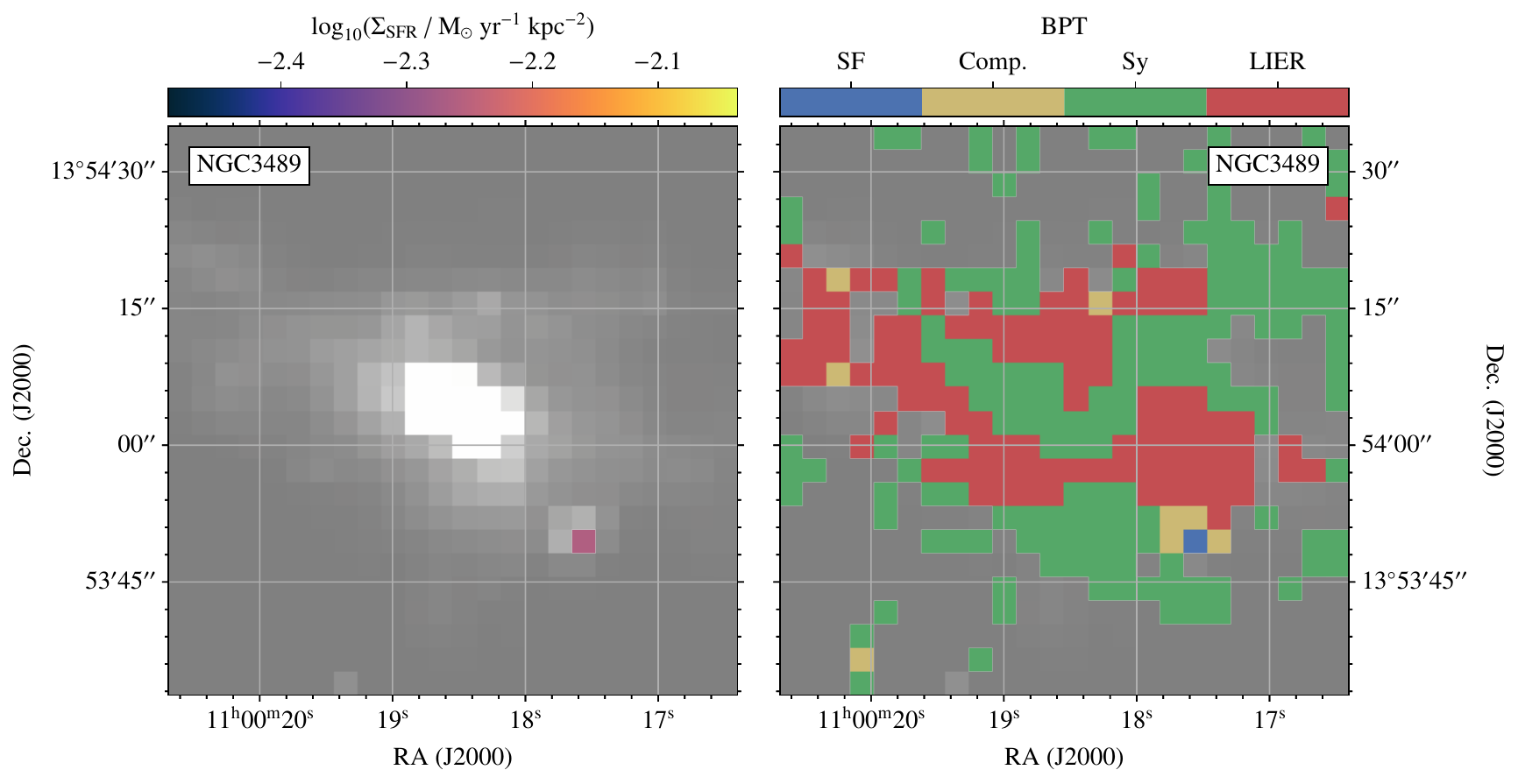}
    \caption{As Figure \ref{fig:ngc4694_sfr_map}, but for NGC~3489.}
    \label{fig:ngc3489_sfr_map}
\end{figure*}

\begin{figure*}
    \includegraphics[width=\textwidth]{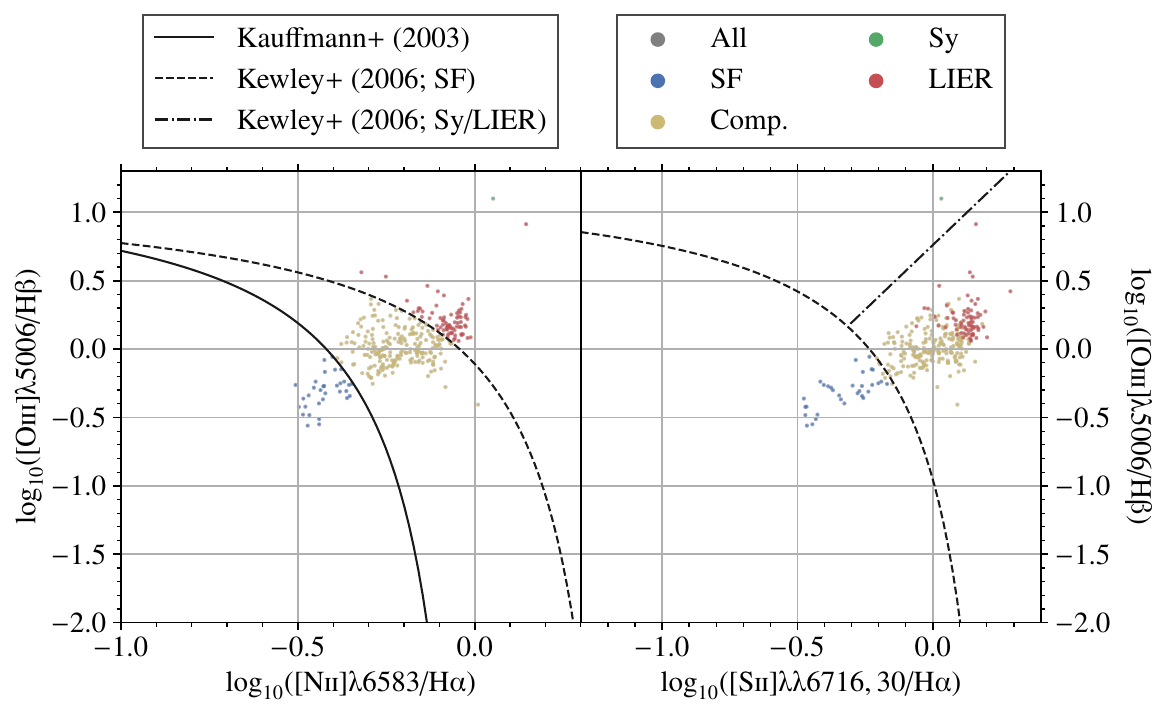}
    \caption{As Figure \ref{fig:ngc4694_bpt}, but for NGC~3599.}
    \label{fig:ngc3599_bpt}
\end{figure*}

\begin{figure*}
    \includegraphics[width=\textwidth]{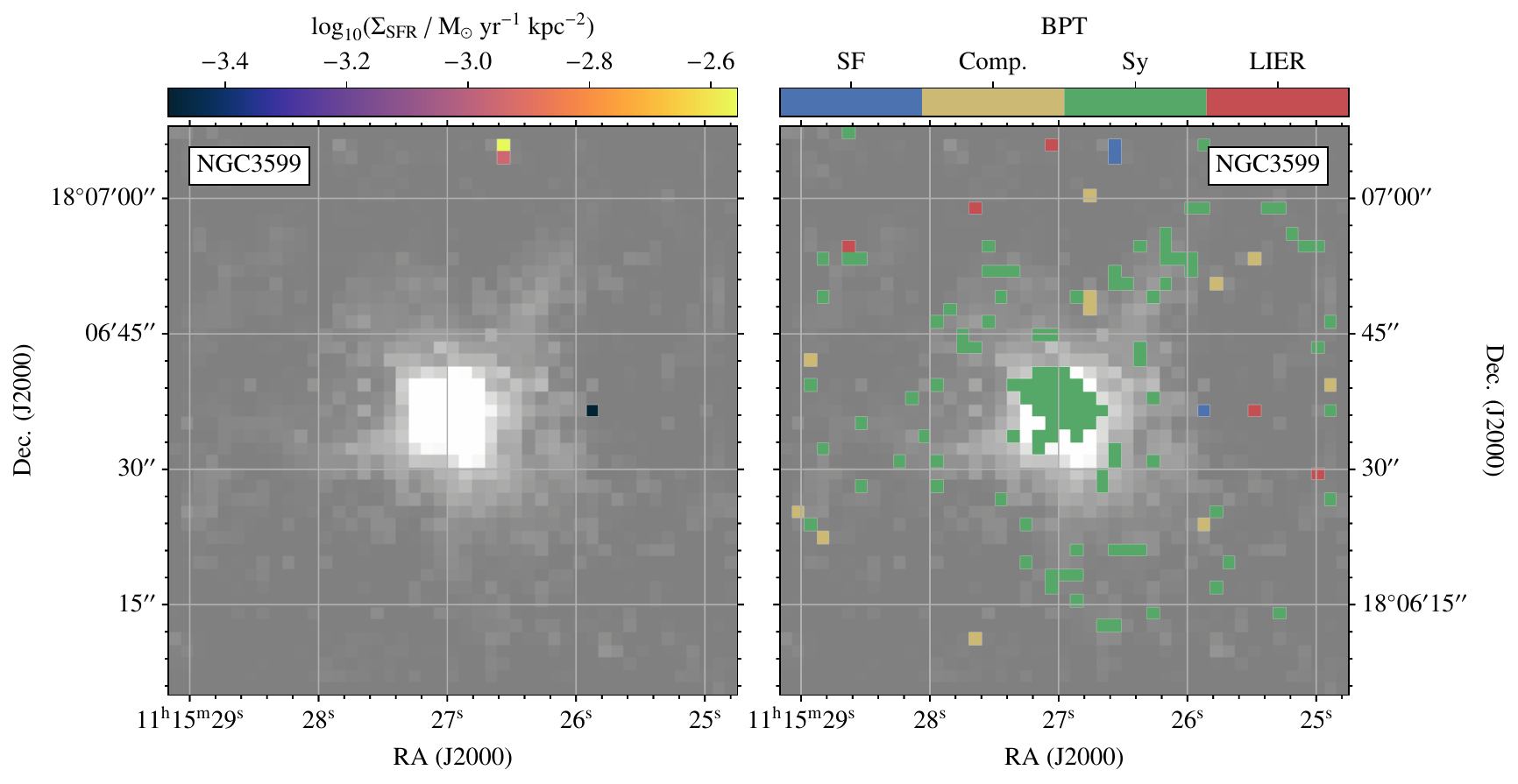}
    \caption{As Figure \ref{fig:ngc4694_sfr_map}, but for NGC~3599.}
    \label{fig:ngc3599_sfr_map}
\end{figure*}

\begin{figure*}
    \includegraphics[width=\textwidth]{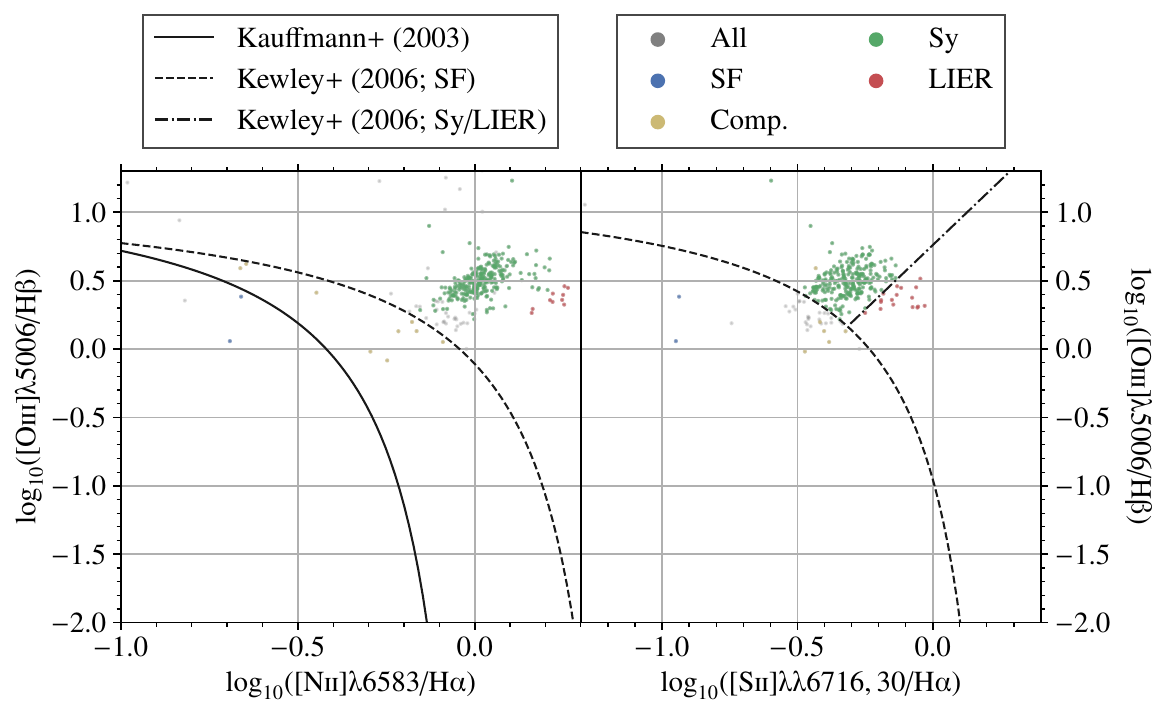}
    \caption{As Figure \ref{fig:ngc4694_bpt}, but for NGC~3607.}
    \label{fig:ngc3607_bpt}
\end{figure*}

\begin{figure*}
    \includegraphics[width=\textwidth]{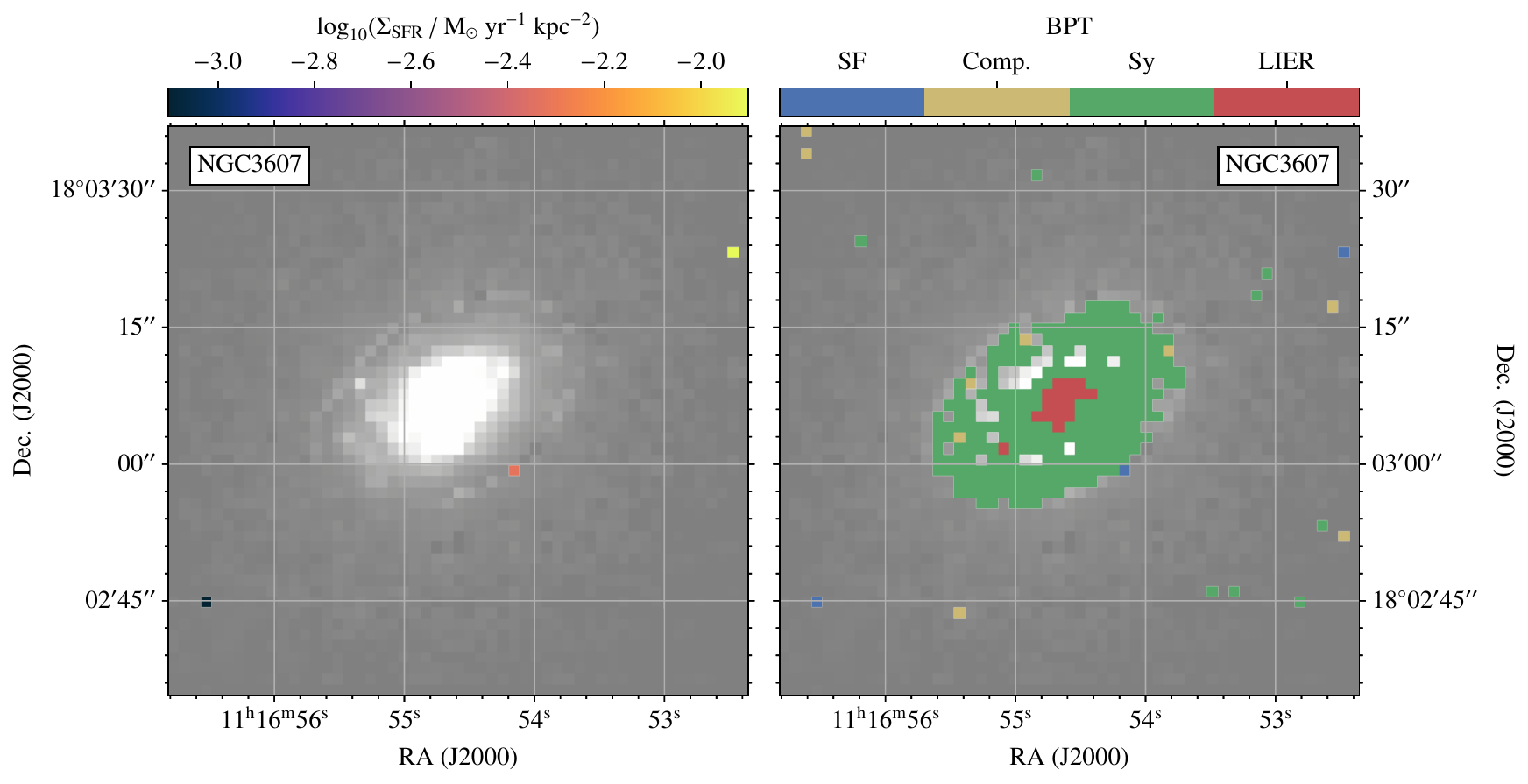}
    \caption{As Figure \ref{fig:ngc4694_sfr_map}, but for NGC~3607.}
    \label{fig:ngc3607_sfr_map}
\end{figure*}

\begin{figure*}
    \includegraphics[width=\textwidth]{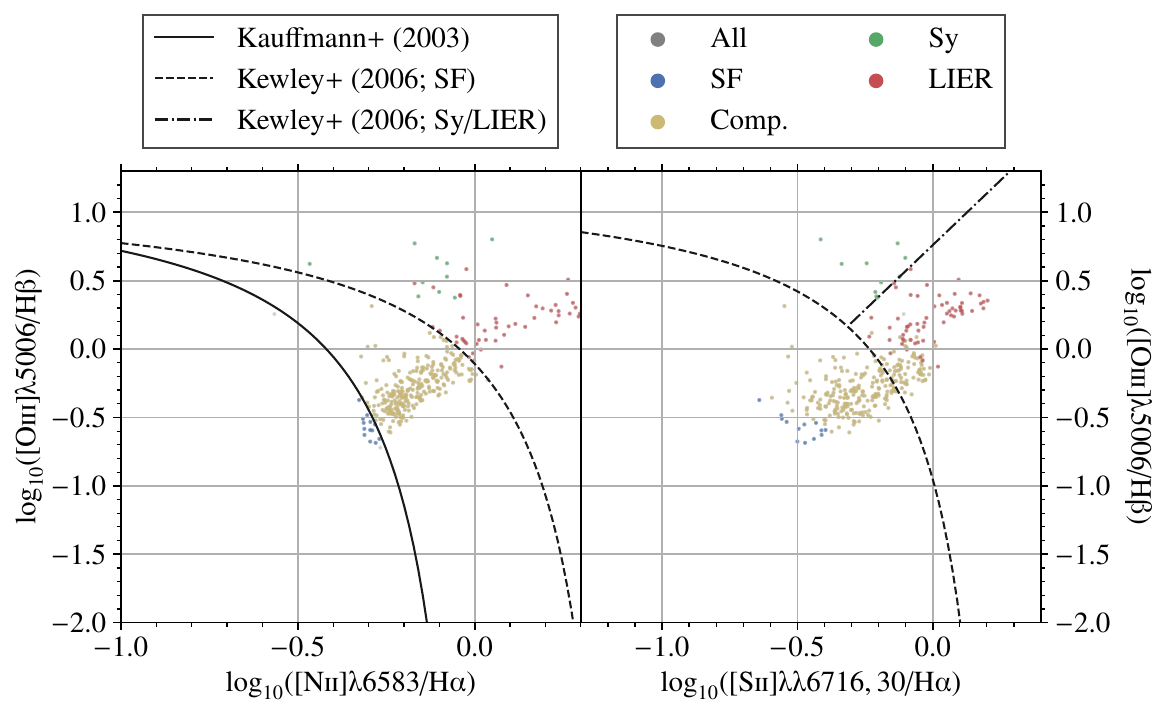}
    \caption{As Figure \ref{fig:ngc4694_bpt}, but for NGC~3626.}
    \label{fig:ngc3626_bpt}
\end{figure*}

\begin{figure*}
    \includegraphics[width=\textwidth]{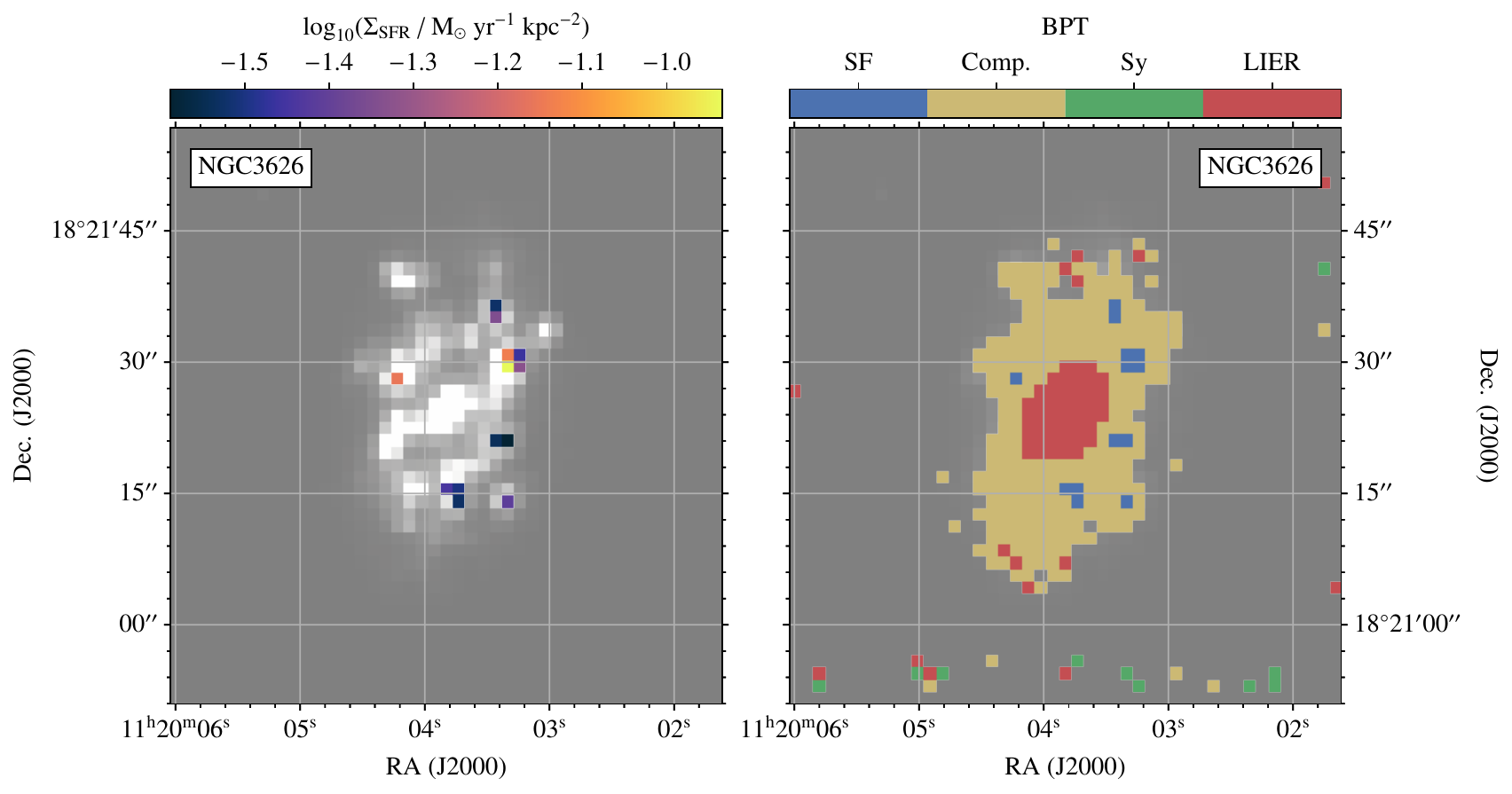}
    \caption{As Figure \ref{fig:ngc4694_sfr_map}, but for NGC~3626.}
    \label{fig:ngc3626_sfr_map}
\end{figure*}

\begin{figure*}
    \includegraphics[width=\textwidth]{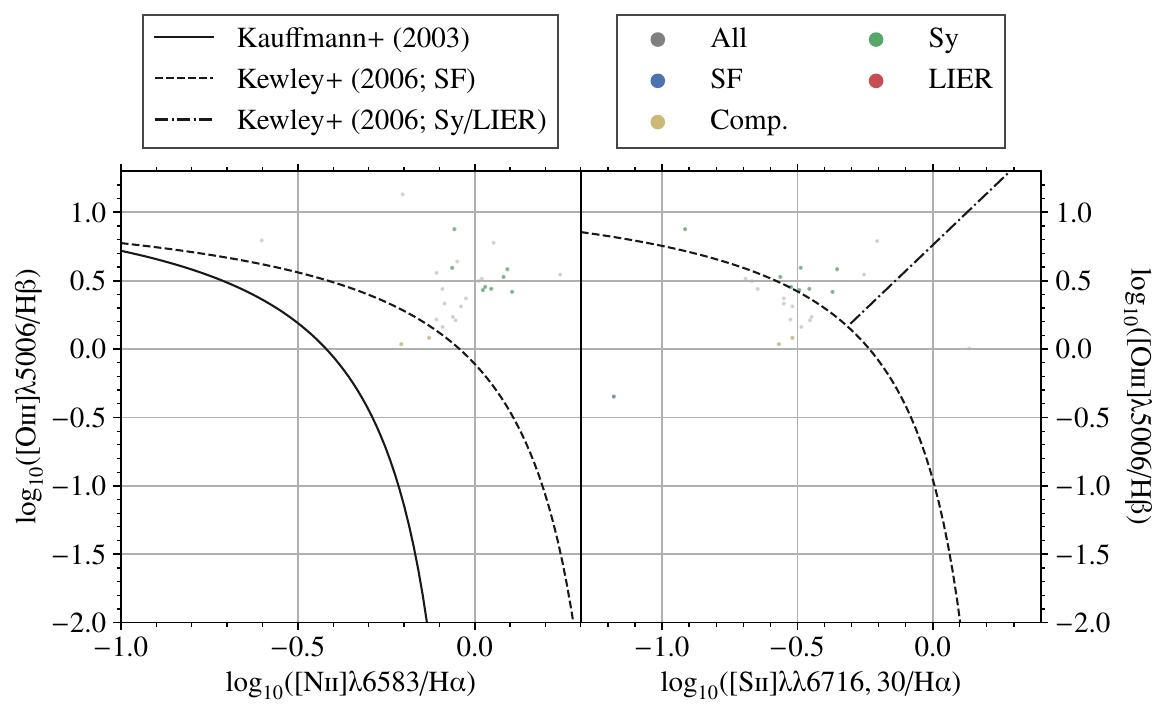}
    \caption{As Figure \ref{fig:ngc4694_bpt}, but for NGC~4435.}
    \label{fig:ngc4435_bpt}
\end{figure*}

\begin{figure*}
    \includegraphics[width=\textwidth]{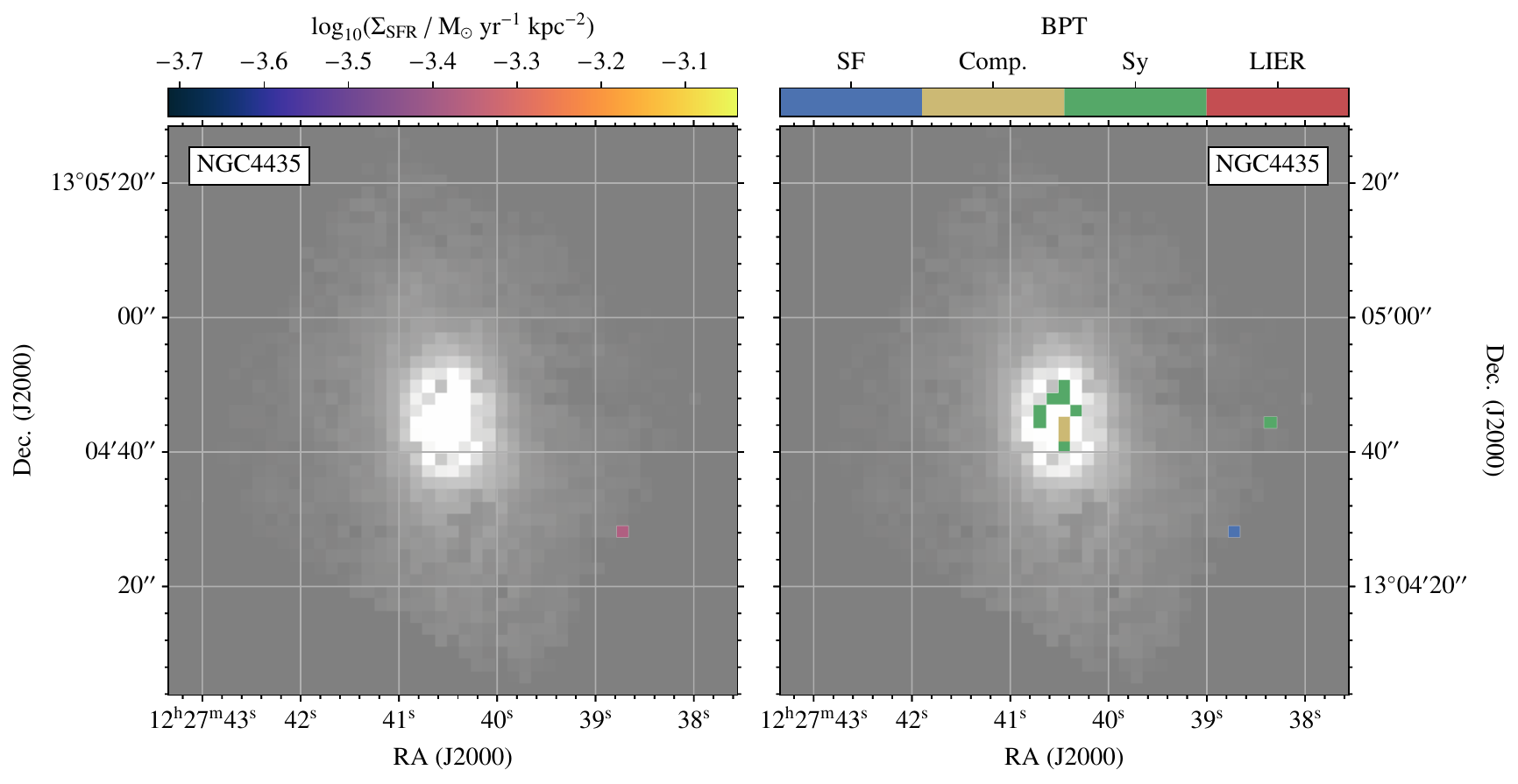}
    \caption{As Figure \ref{fig:ngc4694_sfr_map}, but for NGC~4435.}
    \label{fig:ngc4435_sfr_map}
\end{figure*}

\begin{figure*}
    \includegraphics[width=\textwidth]{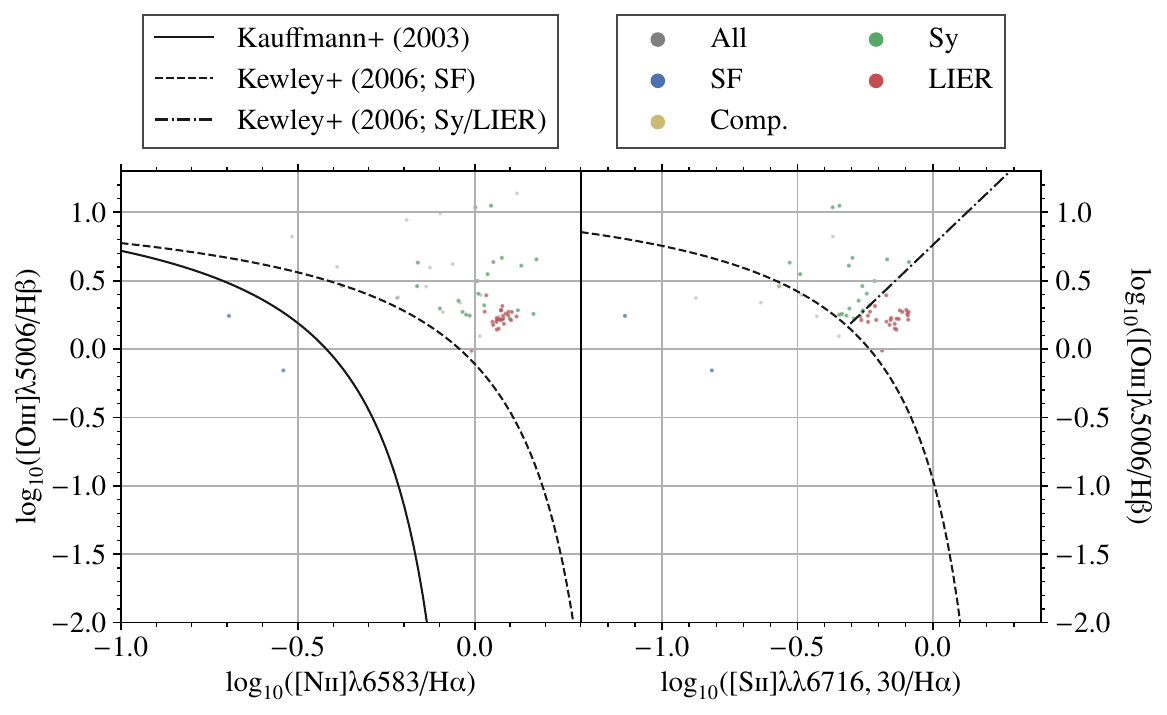}
    \caption{As Figure \ref{fig:ngc4694_bpt}, but for NGC~4596.}
    \label{fig:ngc4596_bpt}
\end{figure*}

\begin{figure*}
    \includegraphics[width=\textwidth]{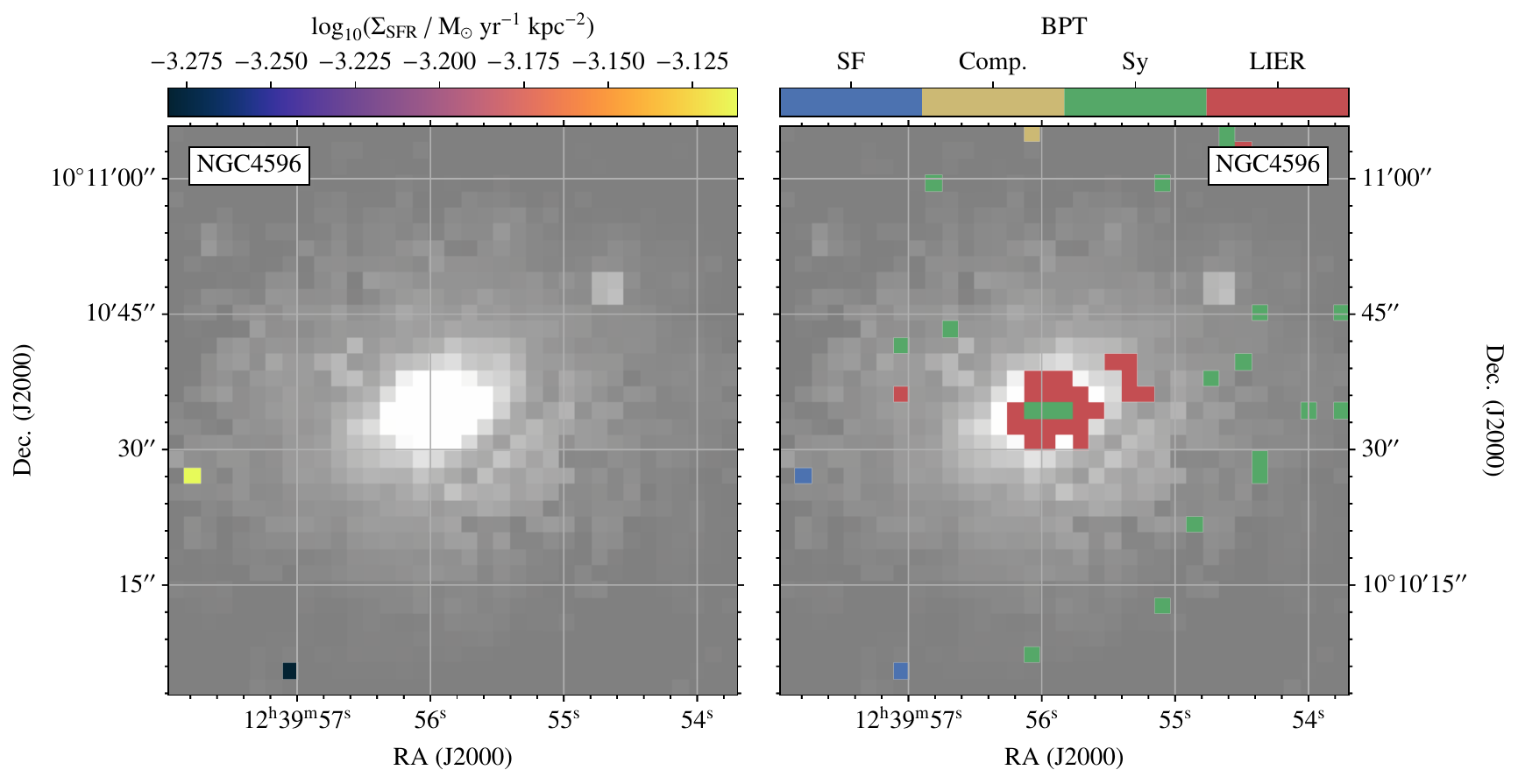}
    \caption{As Figure \ref{fig:ngc4694_sfr_map}, but for NGC~4596.}
    \label{fig:ngc4596_sfr_map}
\end{figure*}

\begin{figure*}
    \includegraphics[width=\textwidth]{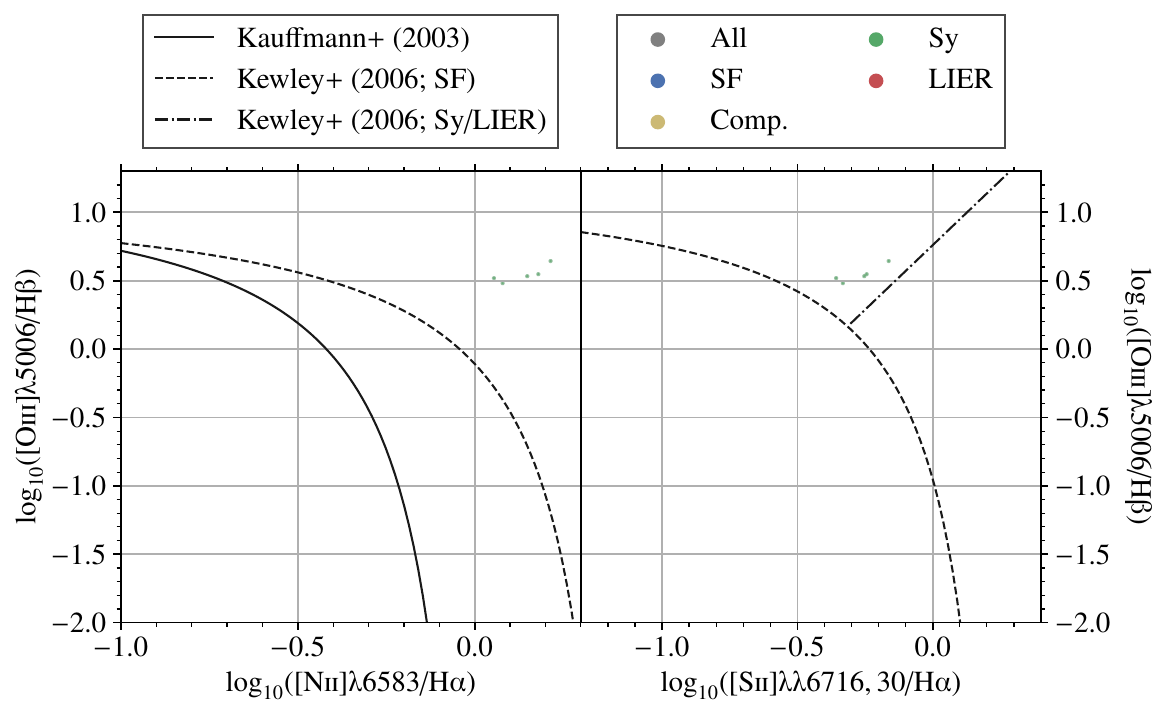}
    \caption{As Figure \ref{fig:ngc4694_bpt}, but for NGC~4697.}
    \label{fig:ngc4697_bpt}
\end{figure*}

\begin{figure*}
    \includegraphics[width=\textwidth]{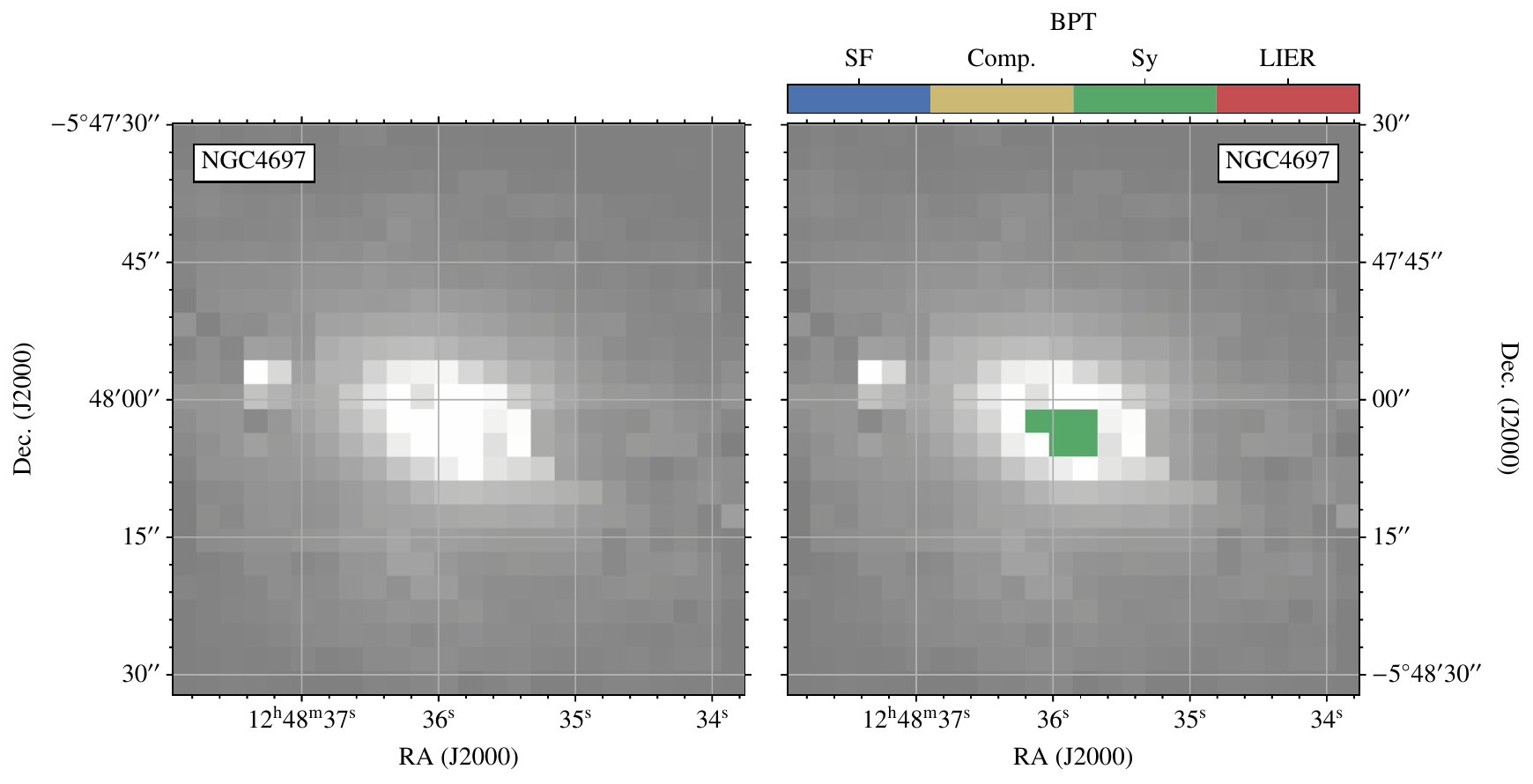}
    \caption{As Figure \ref{fig:ngc4694_sfr_map}, but for NGC~4697.}
    \label{fig:ngc4697_sfr_map}
\end{figure*}

\begin{figure*}
    \includegraphics[width=\textwidth]{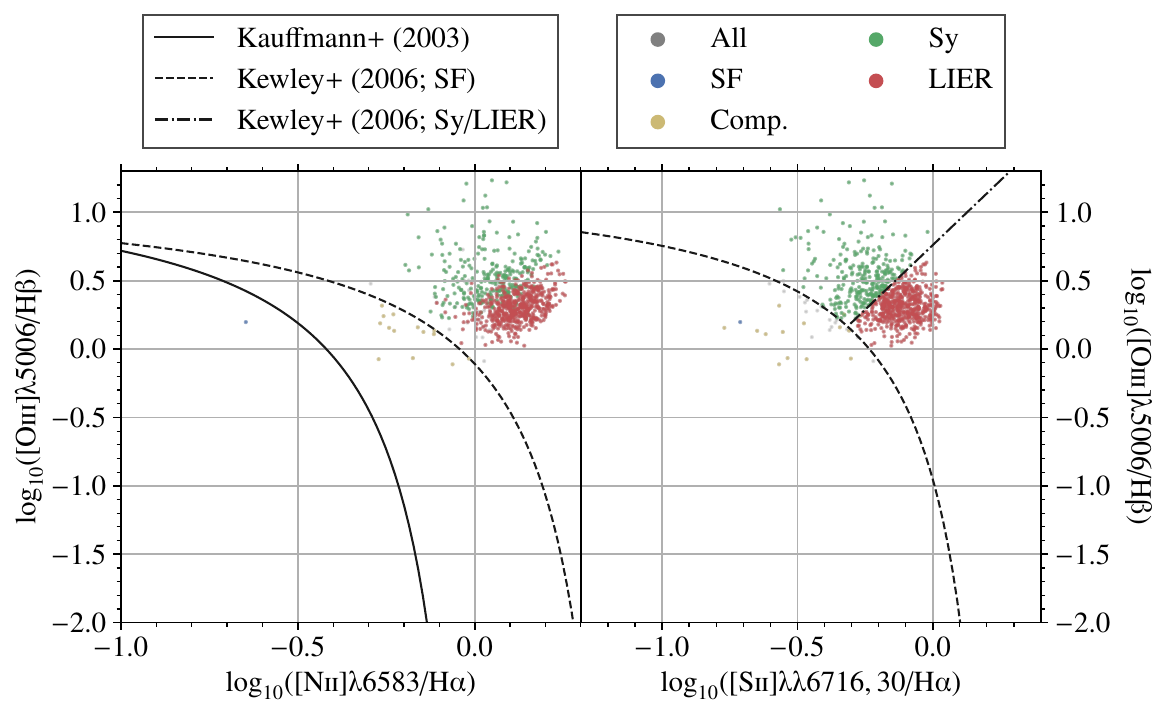}
    \caption{As Figure \ref{fig:ngc4694_bpt}, but for NGC~7743.}
    \label{fig:ngc7743_bpt}
\end{figure*}

\begin{figure*}
    \includegraphics[width=\textwidth]{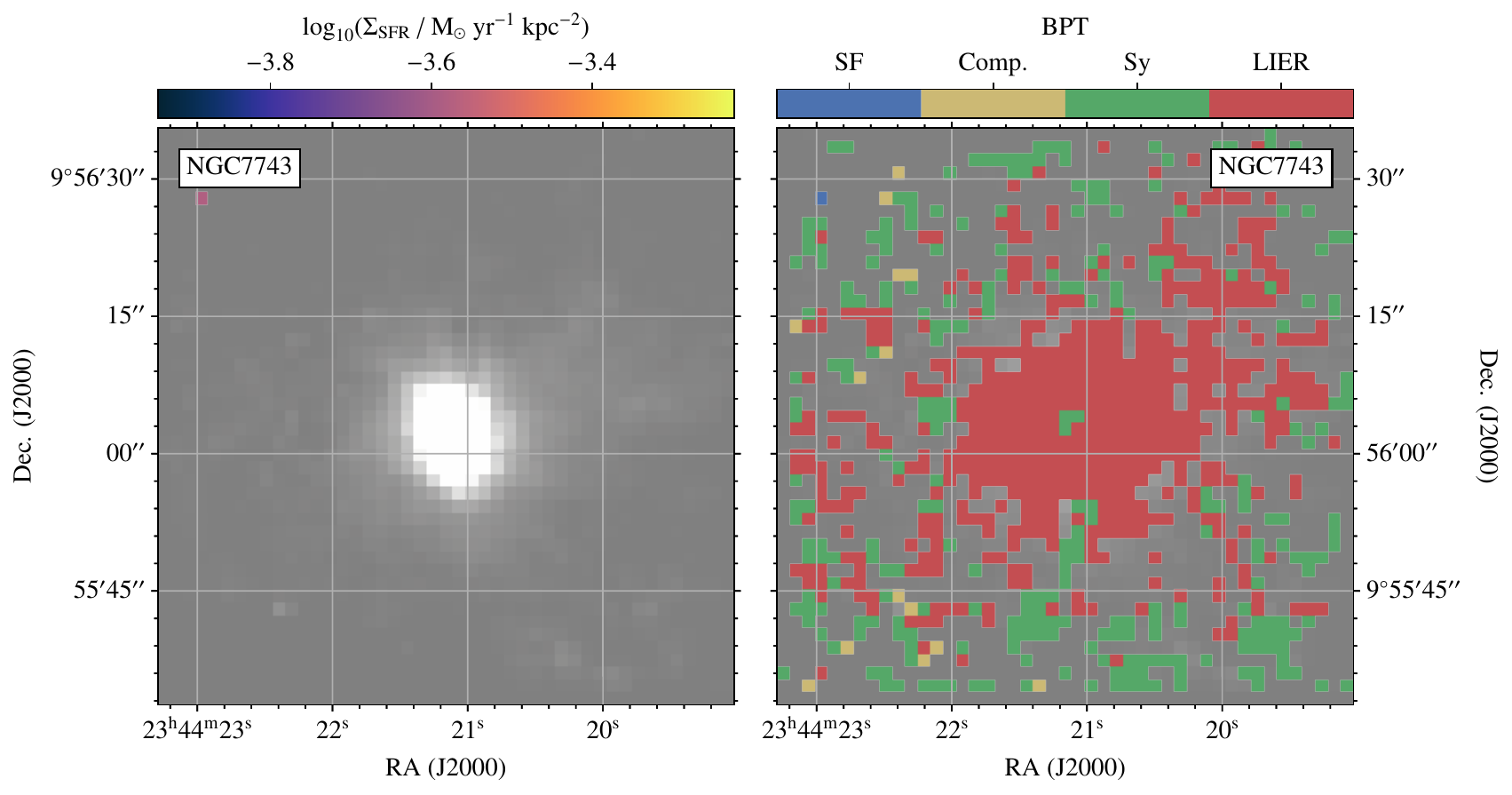}
    \caption{As Figure \ref{fig:ngc4694_sfr_map}, but for NGC~7743.}
    \label{fig:ngc7743_sfr_map}
\end{figure*}

\section{WHAN diagrams for all galaxies}\label{app:all_whans}

Here, we show figures analogous to Figure \ref{fig:ngc4694_whan} for the other galaxies of our sample.

\begin{figure*}
    \includegraphics[width=\textwidth]{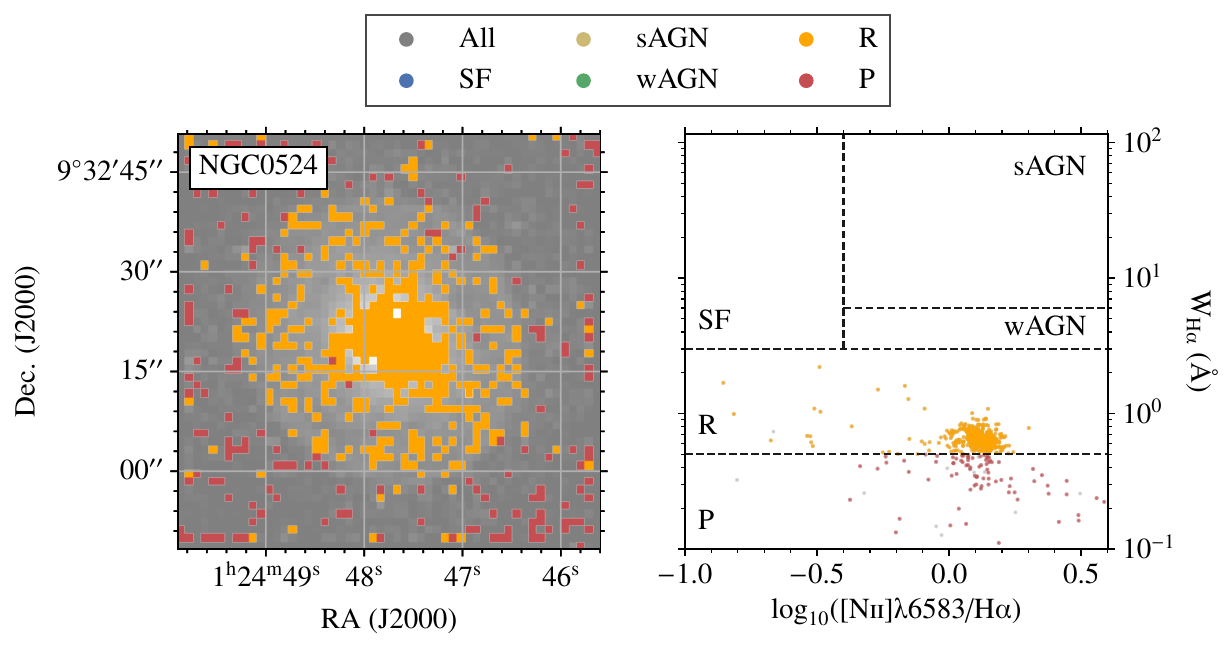}
    \caption{As Figure \ref{fig:ngc4694_whan}, but for NGC~0524.}
    \label{fig:ngc0524_whan}
\end{figure*}

\begin{figure*}
    \includegraphics[width=\textwidth]{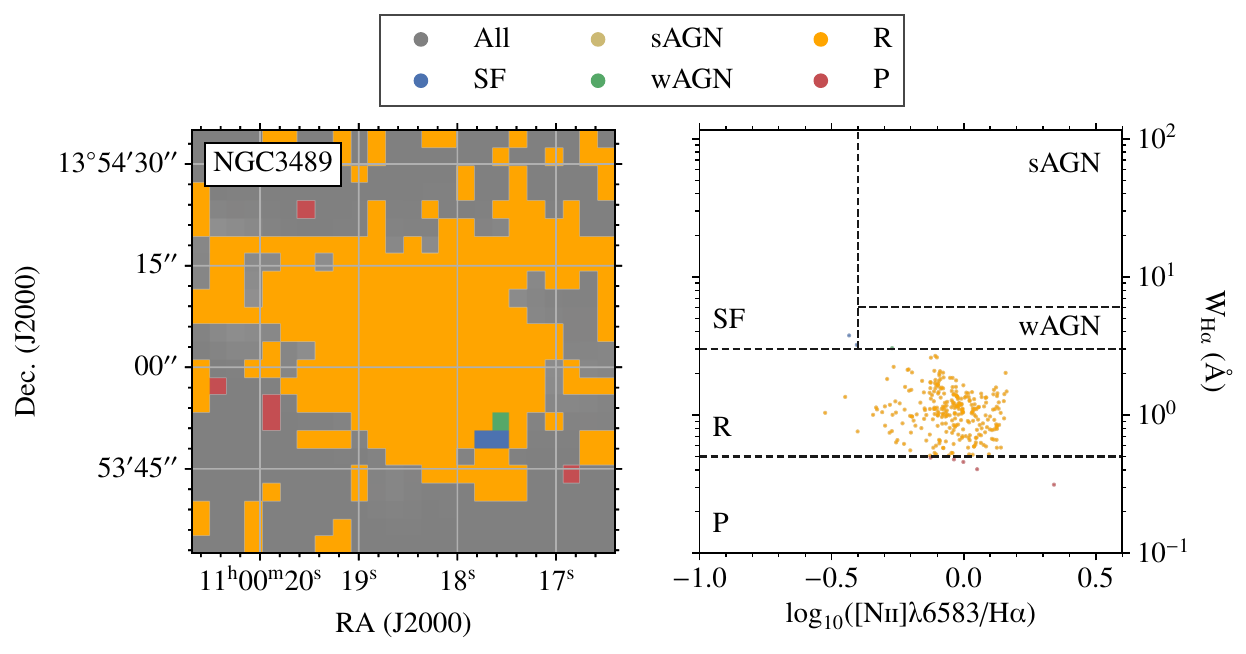}
    \caption{As Figure \ref{fig:ngc4694_whan}, but for NGC~3489.}
    \label{fig:ngc3489_whan}
\end{figure*}

\begin{figure*}
    \includegraphics[width=\textwidth]{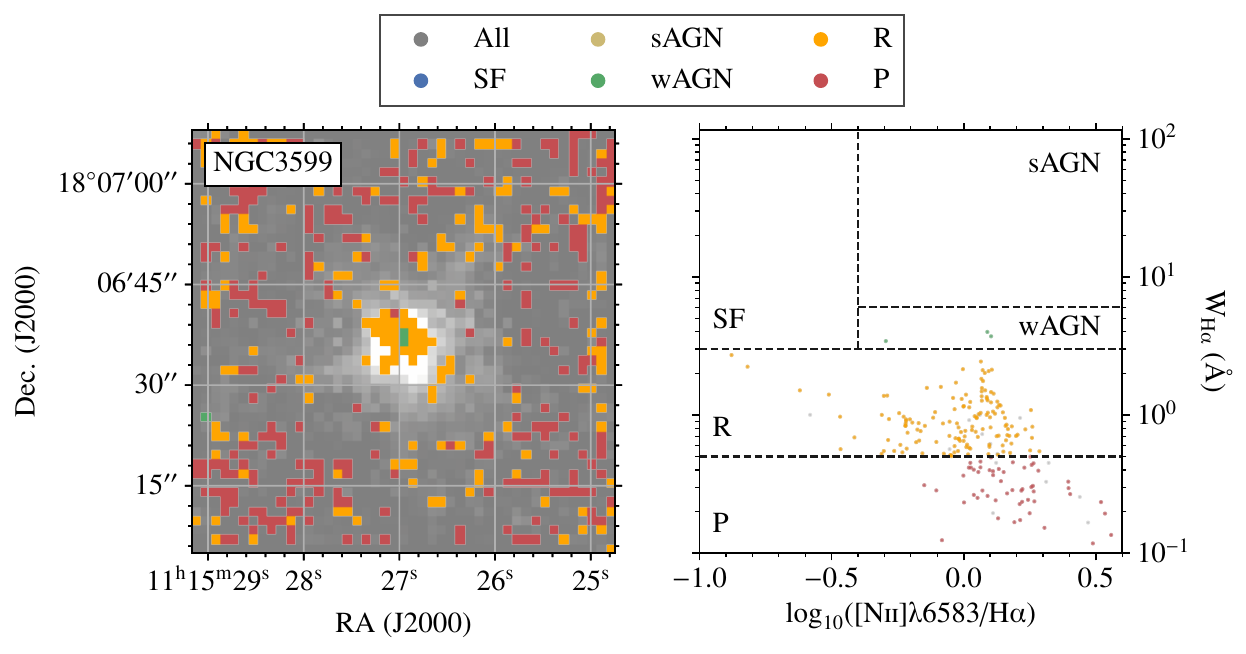}
    \caption{As Figure \ref{fig:ngc4694_whan}, but for NGC~3599.}
    \label{fig:ngc3599_whan}
\end{figure*}

\begin{figure*}
    \includegraphics[width=\textwidth]{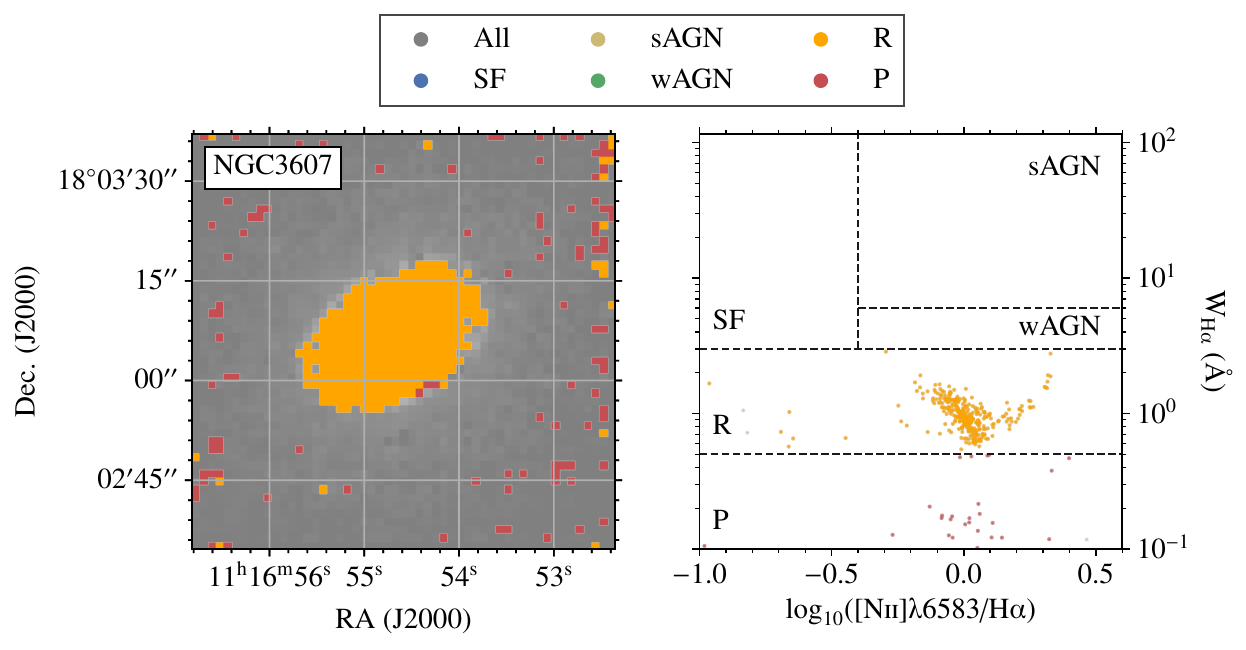}
    \caption{As Figure \ref{fig:ngc4694_whan}, but for NGC~3607.}
    \label{fig:ngc3607_whan}
\end{figure*}

\begin{figure*}
    \includegraphics[width=\textwidth]{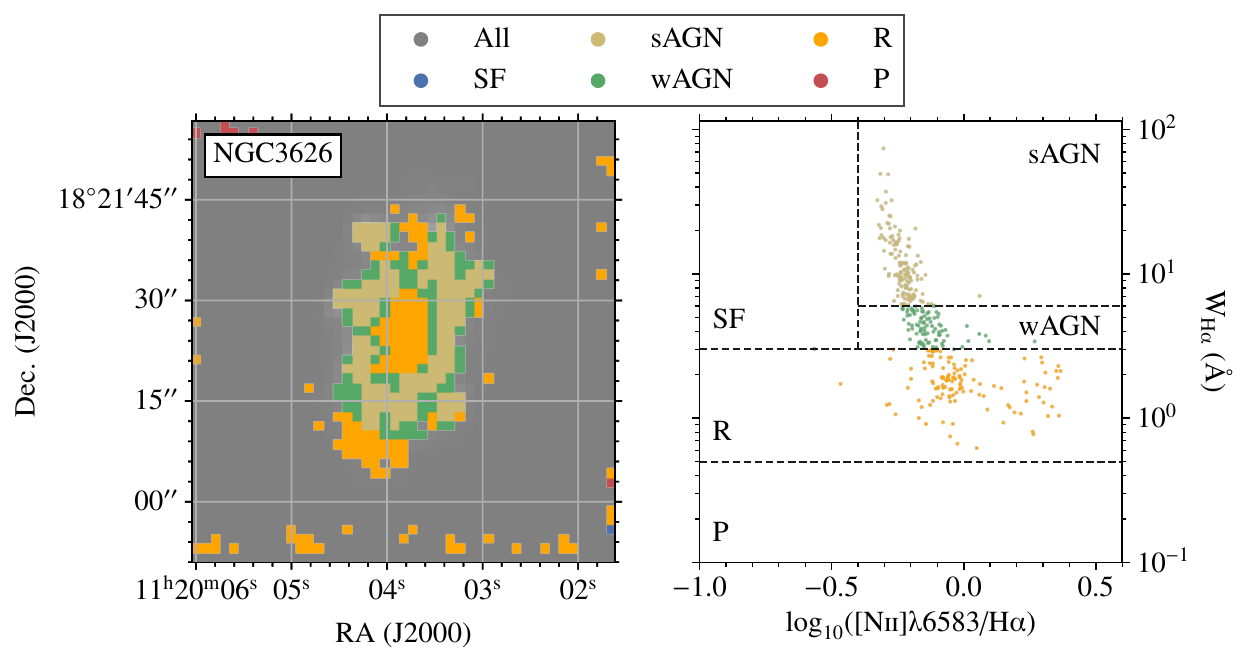}
    \caption{As Figure \ref{fig:ngc4694_whan}, but for NGC~3626.}
    \label{fig:ngc3626_whan}
\end{figure*}

\begin{figure*}
    \includegraphics[width=\textwidth]{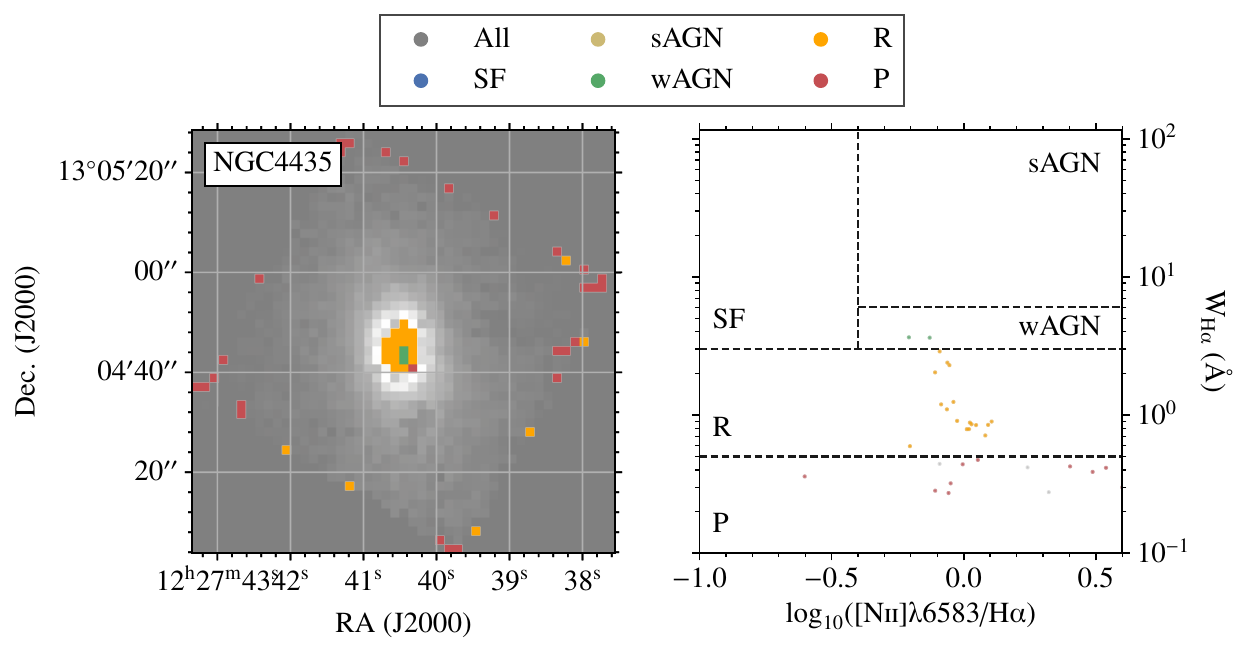}
    \caption{As Figure \ref{fig:ngc4694_whan}, but for NGC~4435.}
    \label{fig:ngc4435_whan}
\end{figure*}

\begin{figure*}
    \includegraphics[width=\textwidth]{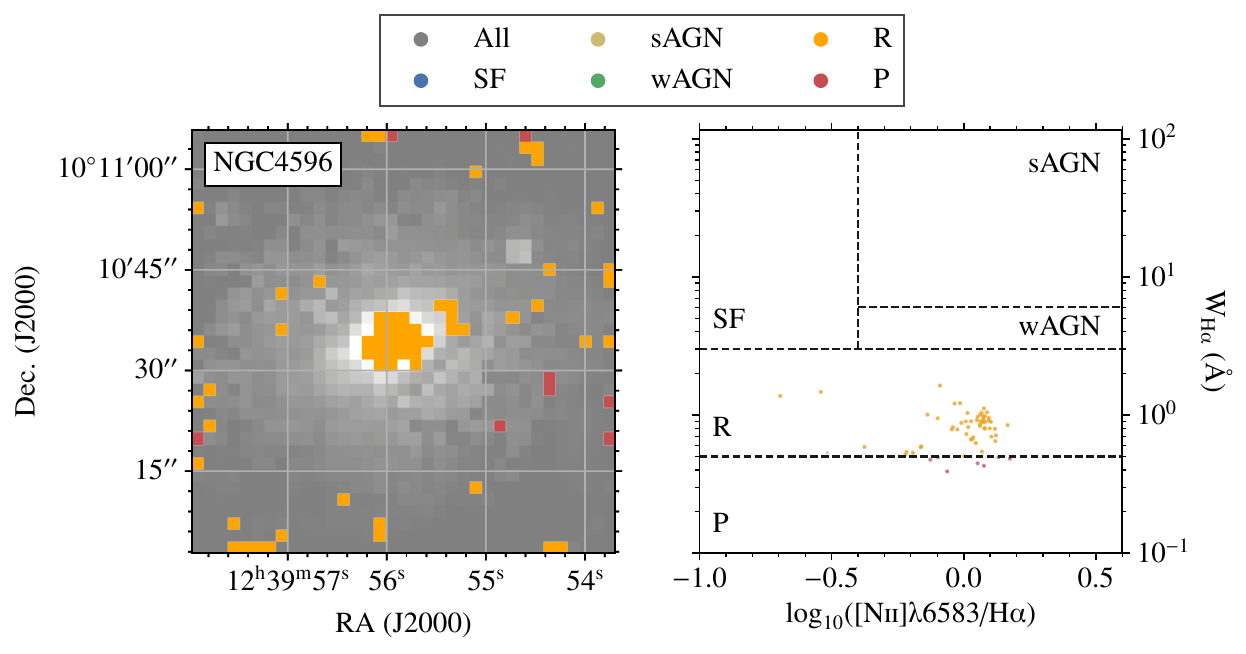}
    \caption{As Figure \ref{fig:ngc4694_whan}, but for NGC~4596.}
    \label{fig:ngc4596_whan}
\end{figure*}

\begin{figure*}
    \includegraphics[width=\textwidth]{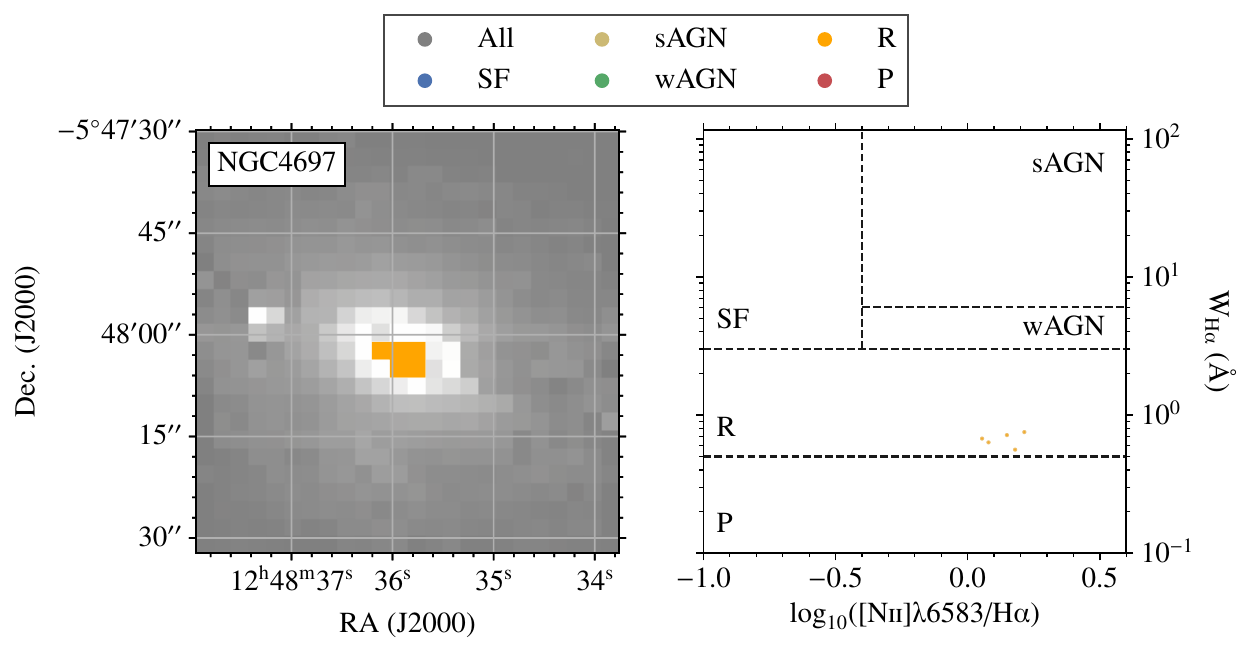}
    \caption{As Figure \ref{fig:ngc4694_whan}, but for NGC~4697.}
    \label{fig:ngc4697_whan}
\end{figure*}

\begin{figure*}
    \includegraphics[width=\textwidth]{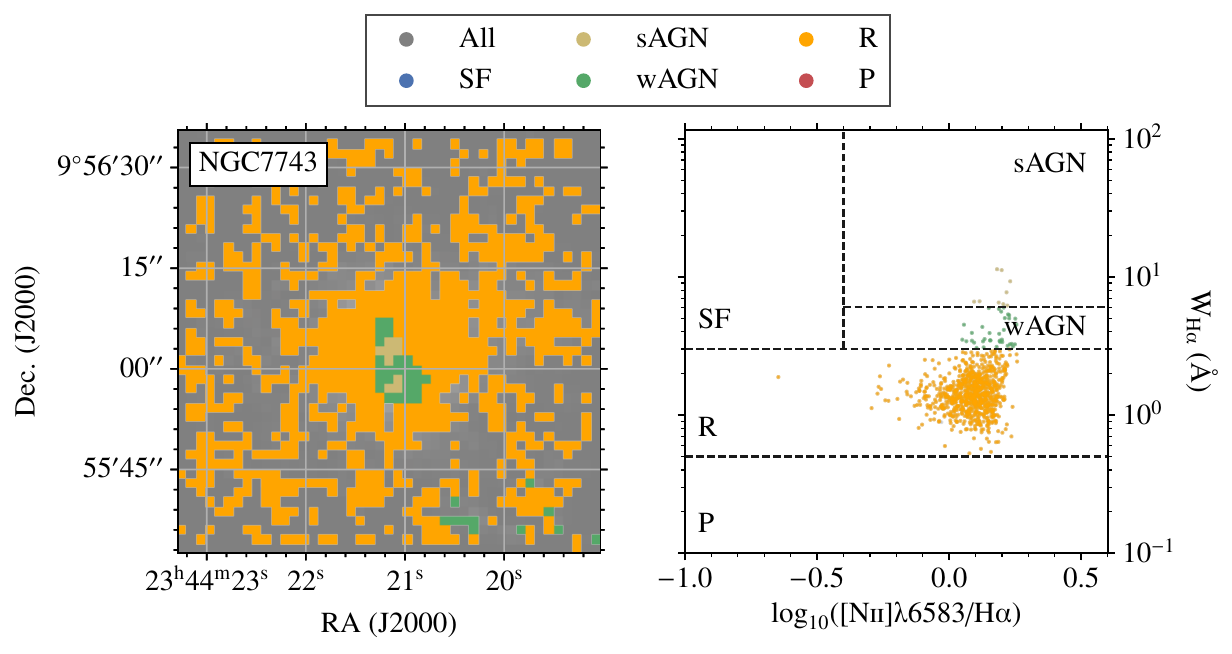}
    \caption{As Figure \ref{fig:ngc4694_whan}, but for NGC~7743.}
    \label{fig:ngc7743_whan}
\end{figure*}


\bsp	
\label{lastpage}
\end{document}